\documentstyle[graphics]{aa} 
\newcommand{\bS}{\mbox{$\mathit{S}$}}

\def\spose#1{\hbox to 0pt{#1\hss}} 
\def\lta{\mathrel{\spose{\lower 3pt\hbox{$\mathchar"218$}} 
        \raise 2.0pt\hbox{$\mathchar"13C$}}}      
\def\gta{\mathrel{\spose{\lower 3pt\hbox{$\mathchar"218$}} 
        \raise 2.0pt\hbox{$\mathchar"13E$}}} 
\def\msol{M$_\odot$}
\begin{document} 
\thesaurus{06	     % A&A Section 6: Form. struct. and evolut. of stars
          (02.01.2;  % Accretion, accretion discs
           02.09.1;  % Instabilities
           08.02.1;  % binaries: close
           13.25.3)} % X-rays: general
\title{The disc instability model for X-ray transients: evidence for truncation and irradiation}
\author{Guillaume Dubus\inst{1}\thanks{Present Address : Theoretical Astrophysics, Caltech 130-33, Pasadena, CA 91125, USA} \and Jean-Marie Hameury\inst{2} \and Jean-Pierre Lasota\inst{3}}
\mail{G. Dubus, gd@tapir.caltech.edu} 
\institute{Astronomical Institute ``Anton Pannekoek'', University of Amsterdam, Kruislaan 403, 1098 SJ, Amsterdam, the Netherlands
\and
Observatoire de Strasbourg, UMR 7550 du CNRS, 11 rue de l'Universit\'e, 67000 Strasbourg, France
\and
Institut d'Astrophysique de Paris, 98bis Boulevard Arago, 75014 Paris, France}
\date{Accepted February 9, 2001}
\titlerunning{Accretion discs in Soft X-ray Transients}
\authorrunning{Dubus, Hameury, Lasota}

\maketitle

\abstract{We study the prospect of explaining the outbursts of Soft
X-ray Transients (SXTs) by the thermal-viscous instability in a thin
disc. This instability is linked to hydrogen ionization and is
significantly changed when irradiation of the disc by X-rays from the
inner regions is included.  We present the first numerically
reliable, physically consistent calculations of the outburst cycles
which include the effects of accretion disc irradiation.  The
decay from outburst is governed by irradiation, as pointed out by King
\& Ritter (1998), leading to slow exponential decays. At the end of
the outburst, the disc is severely depleted, which lengthens the time
needed to rebuild mass to the critical density for an
outburst. Despite this, the long recurrence times and quiescent X-ray
luminosities of SXTs still require the inner disc to be replaced by
another type of flow in quiescence, presumably a hot advection
dominated accretion flow (ADAF). We include the effects of such
truncation and show that the resulting lightcurves are conclusively
similar to those of SXTs like A0620-00. We conclude that the
two-$\alpha$ disc instability model provides an adequate description
of the outbursts of SXTs when both truncation and irradiation are
included. The values for the viscosities in outburst and in quiescence
are comparable to those used in CVs. We discuss the model in the
context of present observations.}

\keywords{Accretion, accretion discs, Instabilities, binaries: close, 
X-rays: general}

\section{Introduction}

Low mass X-ray binaries (LMXBs) are systems in which a low-mass donor
star loses matter to a neutron star or a black hole via Roche lobe
overflow. The infalling gas forms an accretion disc where it diffuses
towards the central body by losing its angular momentum
through viscous interactions. Soft X-ray Transients (SXTs), a
sub-class of LMXBs, show large amplitude X-ray and optical outbursts
during which their luminosity increases by several orders of magnitude
within a few days. The peak is sometimes followed by an exponential
decay on a timescale of a month before the system returns to its very
low luminosity quiescent state. This `fast-rise exponential-decay'
(FRED) behaviour is considered to be typical of SXT lightcurves and is
at the focus of our modeling effort. One should keep in mind,
however, that many if not most of the SXT outbursts do not have this
shape. The recurrence time between outbursts is from several months to
tens of years (see the reviews of Tanaka \& Shibazaki \cite{tanaka}
and Chen et al. \cite{chen}). All of the known black hole
LMXBs are transient systems.

LMXBs and cataclysmic variables (CVs) show many similar
characteristics and the analogy has led to the interpretation of the
outbursts of SXTs as being due to a thermal-viscous instability in a
geometrically thin, optically thick Keplerian accretion disc, just as
in dwarf novae (e.g. Cannizzo et al. \cite{cgw}).  In this disc
instability model (DIM), a steady-state flow is stable if hydrogen is
everywhere ionized. This would be the case for persistent LMXBs and
their CV cousins, the novae-like systems. But if the temperature or
(equivalently) the mass accretion rate from the donor star
$\dot{M}_{\rm tr}$ is low enough for hydrogen to recombine,
the disc becomes thermally and viscously unstable, oscillating between
a hot, ionized state (outburst) and a cold, neutral state
(quiescence). This is due to the strong dependence of the opacity on
temperature when hydrogen is partially ionized (for reviews of the
DIM, see Osaki \cite{osaki96}; Cannizzo \cite{cannizzo2}; Lasota
\cite{review}).

Unfortunately, in this standard form, the DIM fails to provide an
adequate model of LMXBs. For instance, the DIM predicts almost all LMXBs
to be transient systems, in clear contradiction with observations. Van
Paradijs (\cite{vp}) noted that the X-ray flux coming from (or close to)
the compact object could heat the outer accretion disc above the
hydrogen ionization temperature and therefore stabilize the flow. There
is strong observational evidence of such irradiation e.g. the optical
emission of LMXBs which comes mainly from X-rays reprocessed in the
outer disc (van Paradijs \& McClintock \cite{vp2}). Yet, the simple
model consisting of a point source irradiating a thin disc does not lead
to any significant heating of the outer disc (Dubus et al. \cite{dubus};
see also Meyer \& Meyer-Hofmeister \cite{meyer1}). The geometry is
certainly more complex and may involve {\em indirect} rather than
direct irradiation of the disc. A corona reflecting only a small
fraction ($\sim 0.5$\%) of the X-ray luminosity from the center towards
the outer disc would result in enough heating to quench the instability.
Similarly, a disc warped out of the orbital plane could intercept enough
of the central X-ray flux.

Despite this gross uncertainty on the geometry of irradiation heating,
there is no doubt that it plays a major role in the physics of LMXBs.
The obvious next step in finding a model for SXTs, which we address
here, is to examine the consequences of including irradiation heating. 
This is particularly interesting since the standard DIM
completely fails to explain the outbursts of SXTs. Numerical models show
outbursts lasting at most only a few tens of days and with short
recurrence times (Mineshige \& Wheeler \cite{mw}; Menou et al.
\cite{menou1}). Moreover, the standard DIM predicts a rapid decay from
the outburst peak instead of a slower exponential decline. This was
interpreted as a failure of the standard `two-$\alpha$' assumption (see
\S2.4 below) used to describe the viscous interactions in the DIM
(Cannizzo et al. \cite{ccl}). But King \& Ritter (\cite{kr})
pointed out that including irradiation heating might naturally lead to
the observed exponential decays without resorting to more complex
formulations of the viscosity. Previous studies which argued that
irradiation should have a negligible influence on the outburst
properties (e.g. Cannizzo et al. \cite{ccl}, Cannizzo
\cite{cannizzo3}) assumed point source irradiation for which
self-shadowing is crucial (Meyer \& Meyer-Hofmeister \cite{meyer1};
Mineshige et al. \cite{mtw}). We have shown such
arguments to be incorrect for persistent LMXBs and, in all likelihood,
this also applies to SXTs (Dubus et al. \cite{dubus}).

Furthermore, it is quite clear that the thin disc cannot extend down
to the surface of the compact object in quiescence: the X-ray
detection of quiescent SXTs implies mass transfer rates onto the
compact object which are four orders of magnitude higher than those
predicted by the standard DIM (Lasota \cite{lasota1}). This difficulty
is resolved by a two-component accretion flow in which the inner thin
disc gradually evaporates into an advection-dominated accretion flow
(ADAF) during the decline (Lasota et al. \cite{lny} ; for reviews, see
Lasota 1999; Narayan, Mahadevan \& Quataert 1999 ; note that a
different model was suggested for neutron-star SXTs by Brown et
al. \cite{brown}).  This idea has provided a good framework to
understand the spectra of SXTs (e.g. Esin et al. \cite{esin2}), their
quiescent luminosity (e.g. Narayan et al. \cite{narayan2}) and the
X-ray/optical delay during the rise of the outburst (Hameury et
al. \cite{hameury3}). The evaporated disc does not participate in the
thermal-viscous instability and, from a dynamical viewpoint, the disc
can essentially be considered as truncated inside.

The outburst lightcurves predicted by the truncated disc instability
models (TDIM) have been recently investigated by Menou et al.
(\cite{menou1}) ; (see also Cannizzo, \cite{cannizzo3}; Meyer \&
Meyer-Hofmeister, \cite{meyer4}). Menou et al. showed that, if the
quiescent viscosity is smaller (by factors $\sim$ 4) than the value
usually assumed in dwarf novae, the TDIM results are in reasonable
agreement with some of the observed properties of SXTs. Standard
values of viscosity in quiescence lead to outbursts which are too
short,  with reflares (see below \S\ref{p:reflare}), and to short
recurrence times. The major drawback of this model is, however, that
the {\em observed} disc irradiation is not taken into account.

In this work we show that a modified DIM including the effects of
evaporation and irradiation can reproduce reasonably well the
outbursts of SXTs with standard values of the viscosity. The
assumptions and the numerical technique used in the model are
described in \S2. The predictions of the classical DIM are briefly
recalled in \S3. We successively explore the effects of irradiation
heating without evaporation in \S4 and with evaporation in \S5. In \S6
we complete our investigation by varying different parameters of the
full model before discussing the results of this study in \S7.

\section{Numerical model}
The numerical scheme used is that of Hameury et al. (\cite{hameury}),
modified to include irradiation heating as in Dubus et
al. (\cite{dubus}) and Hameury et al. (\cite{hameury2}).
Evaporation of the inner disc is treated in the same way as Menou et
al. (\cite{menou1}). 

Basically, the model takes advantage of the fact that in a thin disc
one can decouple the vertical structure from the radial evolution
equations. The vertical structure equations are solved first,
providing a grid from which the heating and cooling terms can be
obtained during the actual model computation. The radial disc
equations are then solved on an adaptive mesh where the number of grid
points is highest in the regions of {\sl steep} temperature or density
gradients. This results in great accuracy and eliminates pure
numerical uncertainties which plagued earlier attempts (e.g. Mineshige
\& Wheeler \cite{mw}; Cannizzo et al. \cite{ccl}).  We discuss
here some aspects related to the time-dependent irradiated disc model
that were not discussed in the above references.

\subsection{Vertical structure \label{vs}}
The series of disc vertical structures, including irradiation, is
calculated by solving eq. (22--27) of Dubus et al. (\cite{dubus}),
assuming thermal balance and given values for the radius $R$, the
surface column density $\Sigma$, the midplane temperature $T_{\rm c}$
and the irradiation temperature $T_{\rm irr}$. In these equations,
X-ray heating is limited to a thin layer above the disc and appears in
the boundary condition at the surface:
\begin{equation}
T^4_{\rm surf}=T^4_{\rm eff}+T^4_{\rm irr}
\end{equation}
Viscous heating follows the usual
$\alpha$ prescription i.e. is proportional to the local pressure in
the layer $P$ (Shakura \& Sunyaev, \cite{shakura}):
\begin{equation}
{dF_{\rm vis} \over dz}={3\over 2} \alpha \Omega_{\rm K} P
\end{equation}
where $\Omega_{\rm K}=(GM_1/R^3)^{3/2}$ is the Keplerian frequency.

The solutions give for each set of parameters unique values for the
effective temperature $T_{\rm eff}$ and $\alpha$ which are then
stored. We compute 70 $\times$ 120 $\times$ 190 $\times$ 9 vertical
structures in {$\Omega_{\rm K}$, $\Sigma$, $T_{\rm c}$, $T_{\rm irr}$}
($M_1$ and $R$ enter only via the Keplerian frequency).  We typically
consider parameters in the range $10^{8}$--$10^{12}$ cm for $R$,
$10^{-2}$--$10^{3}$ g$\;$cm$^{-2}$ for $\Sigma$, $10^3$--$10^6$ K
for $T_{\rm c}$, and 0--$T_{\rm c}$ for $T_{\rm irr}$. In most cases,
this samples very well the range of values needed by the disc
evolution code.

For given $M_1$, $R$ and $T_{\rm irr}$, the solutions obtained above
show an \bS~shape when plotted in the $\Sigma$--$T_{\rm c}$ plane.
Examples of such irradiated \bS-curves can be found in Tuchman et
al. (\cite{tuchman}) and Dubus et al.  (\cite{dubus}). The low, cold
branch corresponds to a layer of neutral hydrogen while in the upper,
hot branch hydrogen is ionized. In between, the gradual ionization of
hydrogen is associated with a steep increase of the opacity. This
results in the middle branch of the \bS-curve being thermally and
viscously unstable.  A ring of matter on the unstable branch will
evolve either to the hot upper branch or the cold lower branch which
are stable. Neighbouring rings are affected somewhat like dominos and
the instability propagates, periodically switching the disc between
the hot state and cold state. This is the essence of the disc
instability model (DIM) whose application to SXTs we study here.

Following are useful numerical fits to the turning points of the
\bS-curve: $\Sigma_{\rm max}$, $\Sigma_{\rm min}$ and the associated
$T_{\rm c}$ when irradiation is taken into account. $\Sigma_{\rm max}$
is the maximum density on the cold, neutral, branch while $\Sigma_{\rm
min}$ is the minimum density of the hot, ionized, branch. We assume that
$\alpha$ takes a different value on the hot ($\alpha_{\rm h}$) and on
the cold branch ($\alpha_{\rm c}$ ; see below \S\ref{talpha}).
\begin{eqnarray}
\Sigma_{\rm max}=(10.8&-&10.3\xi) \nonumber \\
&\times& \alpha_{\rm c}^{-0.84} M_1^{-0.37+0.1\xi} R_{10}^{1.11-0.27\xi}{\rm ~}\rm{g}\;{\rm cm}^{-2}
\label{simax}
\end{eqnarray}
\begin{equation}
T_{\rm c}(\Sigma_{\rm max})=10700{\rm ~} \alpha_{\rm c}^{-0.1} R_{10}^{-0.05\xi}{\rm ~~K}
\label{tcmax}
\end{equation}
\begin{equation}
\Sigma_{\rm min}=(8.3-7.1\xi){\rm ~} \alpha_{\rm h}^{-0.77} M_1^{-0.37} R_{10}^{1.12-0.23\xi}{\rm ~~}\rm{
g}\;{\rm cm}^{-2}
\label{simin}
\end{equation}
\begin{eqnarray}
T_{\rm c}(\Sigma_{\rm min})=(20900&-&11300\xi){\rm ~}\nonumber \\
&\times&\alpha_{\rm h}^{-0.22}  M_1^{-0.01} R_{10}^{0.05-0.12\xi}{\rm ~K} 
\label{tcmin}
\end{eqnarray}
where $\xi=(T_{\rm irr}/10^4{\rm \ K})^2$, $M_1$ is in solar masses
and $R_{10}=R/10^{10}{\rm cm}$. These fits are very close to those of
Hameury et al. (\cite{hameury}) in the case of no irradiation. Eq.(29)
of Dubus et al. (\cite{dubus}) can be considered as a special case of
Eq.\ref{simin} above for steady-state and a given dependence of
$T_{\rm irr}$ on $R$. At the $\Sigma_{\rm min}$ turning point, the
maximum of the opacity is reached close to the photosphere while it is
at the bottom of the layer at $\Sigma_{\rm max}$ (Cannizzo \& Wheeler,
\cite{canwhe}). Irradiation thus has a much stronger influence on
$T_{\rm c}(\Sigma_{\rm min})$ than on $T_{\rm c}(\Sigma_{\rm max})$.

The \bS-curve disappears when $T_{\rm c}(\Sigma_{\rm max})\geq
T_{\rm c}(\Sigma_{\rm min})$. Irradiation heating is then enough to
fully ionize the layer. This happens for $T_{\rm irr}\approx 10^4$\ K
with a weak dependence on $M_1$, $\alpha$ and $R$ (Tuchman et al.
\cite{tuchman}; Dubus et al. \cite{dubus}). Note that this does not
mean the vertical structure becomes isothermal (Dubus et al.
\cite{dubus}). A disc where $T_{\rm irr}$ is everywhere greater than
$10^4$\ K will be thermally and viscously stable.

\subsection{Disc evolution equations}

The disc evolution is governed by the usual vertically integrated
radial equations for mass, angular momentum and energy conservation:
(Eq. 1-6 of Hameury et al. \cite{hameury} ; see also Cannizzo
\cite{cannizzo}) to which we add an
equation for $T_{\rm irr}$ (Eq.~\ref{ill} below, \S\ref{sectirr}).
In particular, the energy conservation equation is:
\begin{equation}
{\partial T_{\rm c} \over \partial t} = { 2 (Q^+ -Q^- + J) \over C_P \Sigma}
 - {\Re T_{\rm c}
\over \mu C_P} {1 \over r} {\partial (r v_{\rm r}) \over \partial r} -
v_{\rm r} {\partial T_{\rm c} \over \partial r},
\label{eq:heat}
\end{equation}
where $Q^+$ and $Q^-$ are the surface heating and cooling rates
respectively. These are determined using the grid of vertical structures
calculated above. The details of the procedure may be found in \S3 of 
Hameury et al. \cite{hameury}.

\subsection{Irradiation temperature \label{sectirr}}

We have shown in Dubus et al. (\cite{dubus}) that the simple
assumption of a central irradiating point source in the plane 
of a thin disc led to results which were incompatible with
observations. Following previous work (e.g. Shakura \& Sunyaev
\cite{shakura}) we write the irradiation flux as:
\begin{equation}
\sigma T^4_{\rm irr} = {\cal C} \frac{L_{\sc x}}{4 \pi R^2} {\rm
~~with~} L_{\sc x}= \epsilon\ {\rm min\ }(\dot{M}_{\rm
in},\dot{M}_{\rm Edd})\ c^2
\label{ill}
\end{equation}
$\dot{M}_{\rm in}$ is the mass accretion rate at the inner disc radius
and $\dot{M}_{\rm Edd}$ is the Eddington mass accretion rate for an
accretion efficiency $\epsilon$ of 0.1: $\dot{M}_{\rm
Edd}=1.4\times 10^{18}M_1{\rm\ gs}^{-1}$.  The maximum luminosity that
can irradiate the disc is thus Eddington-limited.  In the models
including evaporation (see \S\ref{sec:evap}) $\epsilon$ varies to
reflect the very low efficiency of the ADAF. Without evaporation,
$\epsilon$ is constant, equal to 0.1.

$\cal C$ is a measure of the fraction of the X-ray luminosity that
heats up the disc and, as such, contains information on the
irradiation geometry, X-ray albedo, X-ray spectrum, etc. Kim et al.
(\cite{kim}) used an analogous prescription for their indirect
irradiation flux. We found that the observed optical magnitudes and
stability properties of persistent low mass X-ray binaries were
compatible with a value of $\cal C$$\approx$ 5$\times 10^{-3}$ (with
$\epsilon=0.1$, Dubus et al. \cite{dubus}). Esin and co-workers found
some evidence for changes of $\cal C$ in the outbursts of \object{GRO
J1655-40} (a.k.a. \object{V1033 Sco}), \object{A0620-00}
(a.k.a. \object{V616 Mon}) and \object{GRS 1124-68} (a.k.a. \object{GU
Mus}) (Esin et al. \cite{esin}; \cite{esin3}).  Radiation-driven
warping of the outer disc in long period SXTs would produce a varying
value of $\cal C$ (e.g. Ogilvie \& Dubus, \cite{ogilvie}). For
simplicity, we have nonetheless restricted ourselves to $\cal C$
constant in this study.

\subsection{Temperature dependence for $\alpha$ \label{talpha}}

As in our previous works, we assume that $\alpha$ takes on two
different values on the hot and cold branches of the
\bS-curve. Using a single value for $\alpha$ produces only low
amplitude outbursts as numerous previous works have shown.

Since irradiation modifies the hot branch, we adopt a slightly
different prescription for the dependence of $\alpha$ on temperature
than that used by Hameury et al. (\cite{hameury}; \cite{hameury2}):
\begin{eqnarray} 
\log (\alpha)=\log(\alpha_{\rm c})& +& \left[ \log(\alpha_{\rm h})-
\log( \alpha_{\rm c} ) \right] \nonumber \\ & &
\times \left[1+ \left( \frac{T_{\rm crit}}{T_{\rm c}} \right)^{8}\right]^{-1}
\end{eqnarray} 
where $T_{\rm crit}=0.5\ [T_{\rm c}(\Sigma_{\rm max})+T_{\rm
c}(\Sigma_{\rm min})]$ is calculated using Eq.~\ref{tcmax} and
\ref{tcmin}. This ensures that, even with irradiation,
$\alpha=\alpha_{\rm h}$ on the hot branch and $\alpha=\alpha_{\rm
c}$ on the cold branch.

\subsection{Boundary conditions}
The boundary conditions are those of Hameury et al. (\cite{hameury}).
We stress that in all models, the outer disc radius $R_{\rm out}$
is variable. The removal of angular momentum from the disc by the
tidal forces from the secondary (as expressed in Hameury et al.
\cite{hameury}) sets $R_{\rm out}$. The mass transfer rate
$\dot{M}_{\rm tr}$ from the secondary to the accretion disc appears in
the outer boundary conditions.

The inner boundary conditions change when evaporation is included and
this is discussed below in \S4. In models without evaporation, the
inner disc radius is held fixed at a given $R_{\rm in}$. In our
calculations we chose larger values than appropriate if the disc were
to extend down to the compact object e.g. to $R_{\rm in}=3 R_S$ for a
black hole primary. This is essentially a numerical limitation due
to the very low densities at low radii when the disc enters quiescence
(see Fig.~\ref{fig:decayidim}). This is not a drawback: as stressed
in the introduction, there are strong arguments in favour of a truncated
disc in quiescence rather than a disc extending down to the compact object.

In outburst $R_{\rm in}$ could be expected to be close to the compact
object but our assumption does not change any of the results. In
outburst the inner disc is hot and the viscous timescale at which
density perturbations evolve is very short at low radii. Adjustments
of the inner disc are therefore instantaneous compared to the
evolution timescale as long as $R_{\rm in}$ is low enough. We have
verified this by computing some models with $R_{\rm in}=10^7-10^9$~cm
which showed identical outburst lightcurves. A large $R_{\rm in}$
also avoids the Lightman-Eardley instability which arises when the
radiation pressure dominates in the inner disc. Note however that
strong irradiation can suppress this instability 
(Czerny et al. \cite{ccg}; Mineshige \& Kusunose \cite{mk}).

\subsection{Numerical considerations}

The adaptive mesh numerically resolves the heat fronts and this is one
of the major advantages of this scheme (Hameury et al.
\cite{hameury}). The disc mass is not necessarily conserved in the
method and therefore provides a test of the numerical accuracy. At
least 1200 mesh points were used in the runs shown here, resulting in
excellent mass conservation for all cases. The only previous
attempt at including irradiation heating in the DIM used at most 21
radial zones, casting very serious doubts on the validity of their
results (Mineshige et al. \cite{mtw}; Kim et al. \cite{kim}). Hameury
et al. (\cite{hameury}) have emphasized the importance of having
numerically resolved heat fronts (i.e. requiring much more than 21
grid points) to obtain reliable results. In another study,
Cannizzo (1998) calculated the expected irradiation temperature
throughout the eruption cycle of an {\em unirradiated} disc but not
its feedback on the evolution.

As starting points to our calculations we used the (unstable)
steady-state solutions. We then let the disc relax to the
periodic outburst-quiescence cycle which is independent of the initial
starting point (see \S4.7 of Menou et al. \cite{menou2}).

\section{Non-irradiated discs}

The evolution of the inner mass accretion rate times the efficiency
gives a measure of the bolometric lightcurve of the system ($L_{\rm x}
\approx \epsilon \dot{M}_{\rm in} c^2$) (one should in principle
convolve this with the detector spectral response to compare with the
measured X-ray light curve). In the following, we usually show only
the evolution of $\dot{M}_{\rm in}$, so these are not strictly
speaking lightcurves, even though they have the same general
characteristics.  We show some typical `lightcurves' predicted by the
standard DIM for parameters appropriate to soft X-ray transients
before turning to models including irradiation.

\begin{figure} 
% makefig1.pro
\resizebox{\hsize}{!}{\includegraphics{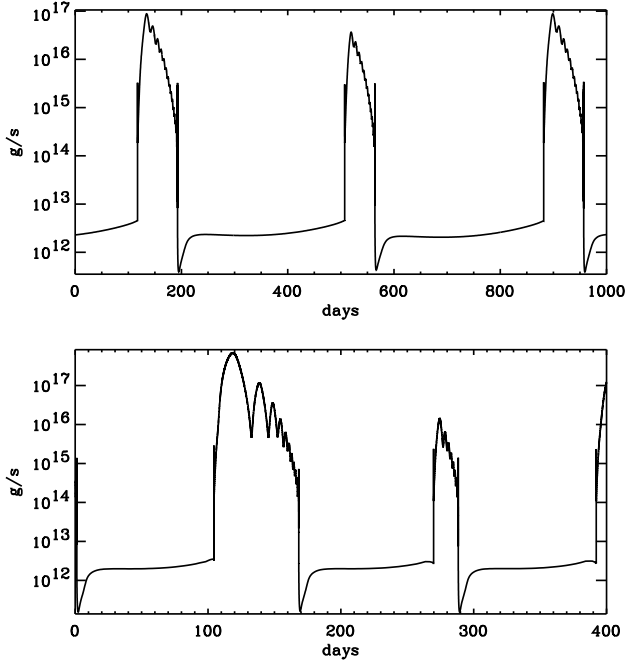}}
\caption{Typical SXT variations of the inner accretion rate predicted
by the standard DIM. This shows the mass accretion at the inner edge
of the disc around a 6~\msol\ black hole (top) and a 1.4~\msol\
neutron star (bottom). The disc is not irradiated and the inner radius
is kept fixed at $R_{\rm in}=5\times10^8$ cm (top) and $2.5\times10^8$
cm (bottom). The other parameters are those of the `A0620-00' black
hole and `Aql X-1' neutron star models of Menou et
al. (\cite{menou1}). The standard DIM applied to SXTs is characterized
by short recurrence times, low accreted mass during outburst and the
presence of strong reflares during the decline. Clearly, this is does
not reproduce observations.
\label{fig:sdim}} 
\end{figure} 

\subsection{Reflares\label{p:reflare}}
\begin{figure}
% makefigreflare sur hermitage radial4
\resizebox{\hsize}{!}{\includegraphics{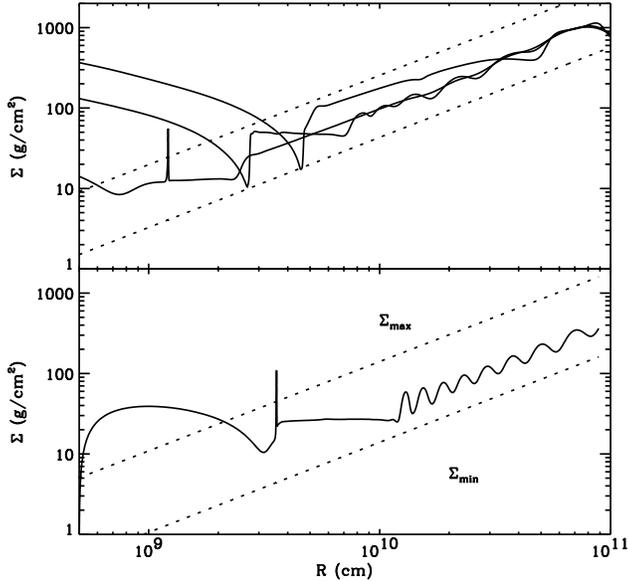}}
\caption{Reflares in non-irradiated discs around a NS (top panel) and
a BH (bottom panel). The full lines are the surface density $\Sigma$
at different times during the decay from outburst
(Fig.~\ref{fig:sdim}) and the dotted lines are $\Sigma_{\rm max}$ and
$\Sigma_{\rm min}$ (Eq.~\ref{simax}, \ref{simin}).  In the top panel,
the situation before the onset of the first reflare is shown by the
top curve: $\Sigma$ just behind the cooling front is close to the
critical $\Sigma_{\rm max}$ for which the disc switches back to the
hot state. The middle and bottom lines show the profile after several
reflares: $\Sigma$ in the outer disc has been depleted by the
successive passages of the heat front (which can be seen in the bottom
curve). The lower panel shows the situation after many successive
reflares in a disc around a BH primary. The wiggles in the density
profile corresponding to previous reflares have not yet been smoothed
out by viscous diffusion. The inner edge is at $10^8$~cm (top) and
$5\times 10^8$~cm (bottom).}
\label{fig:reflare} 
\end{figure}

A characteristic feature of the DIM applied to SXTs are the multiple
reflares that occur during the outburst decay
(Fig.~\ref{fig:sdim}).  These were previously discussed by Menou
et al. (\cite{menou1}). Reflares appear when the surface density
$\Sigma$ behind the cooling front is high enough to reach $\Sigma_{\rm
max}$. At the radius at which this happens, the disc becomes thermally
unstable and a new heating front develops. This front propagates
outwards like an inside-out outburst, reheating the disc until
$\Sigma(R)\la\Sigma_{\rm min}$, when cooling can resume. The density in
the cold region is depleted as matter is accreted during this process,
and the following reflare occurs at smaller radii and have lower
amplitudes (Fig.~\ref{fig:reflare}).

Menou et al. (\cite{menou2}) found the density behind the cooling
front is roughly $K\Sigma_{\rm min}$ where $K$ is a function of the
viscosity $\alpha$ and of the primary mass $M_1$ (Vishniac
\cite{vishniac}). For $M_1=7$ M$_{\odot}$, $K\approx6-7$ so that
$K\Sigma_{\rm min}$ can become greater than $\Sigma_{\rm max}$ and
trigger a heating front. $K$ is lower for lower primary masses
($K\approx4$ for $M_1=1.2$ M$_\odot$) making it less likely to have
reflares. Menou et al. noted reflares were absent in their
calculations with a neutron star primary.

Reflares also appear for neutron stars, but at smaller radii than
considered by Menou et al. The reason is that the post-front $\Sigma$
is not strictly proportional to $\Sigma_{\rm min}$ and has a slightly
shallower dependence on radius: $\Sigma\propto R^{0.9}$ for a NS
(Fig.~\ref{fig:reflare}) instead of $R^{1.1}$ for $\Sigma_{\rm min}$
and $\Sigma_{\rm max}$ (Eq.~\ref{simax}-\ref{simin} without irradiation).
Hence there is always a radius for which $\Sigma$ behind the cooling
front will become greater than $\Sigma_{\rm max}$ and ignite a
reflare. For a neutron star, the first reflare appears after the
cooling front has propagated well inside the disc (Menou et al.
\cite{menou1}, stopped their calculations before this radius was
reached) and has a strong effect on the lightcurve as it heats back a
significant fraction of the disc (bottom panel of Fig.~\ref{fig:sdim}).

\subsection{A deficiency of the DIM}

Reflares are a generic feature of the DIM but do not correspond to
observed features in the decay lightcurves of SXTs (Chen et al.
\cite{chen}). Specifically, the rise-times are much too long in the
calculated reflares. The decreasing amplitudes and timescales
in-between outbursts is different from the regular mini-outburst
displayed by \object{GRO J0422+32} (a.k.a \object{V518 Per}).  The
number of reflares predicted is also incompatible with the glitches of
e.g. \object{A0620-00}.

Reflares are therefore a deficiency of the DIM (Menou et al.
\cite{menou1}).  Models of dwarf novae are not affected by this
problem because the radii of the white dwarf primary ($\sim 10^{9}$
cm) and of the discs (few times $\sim 10^{10}$ cm) do not span a range
large enough for reflares to appear.  The reflares can be avoided in
SXTs if, for instance, the thin disc does not extend much further in
than the radius at which reflares ignite. However, a thin disc
truncated at too high a radius will be cold and stable rather than
transient.

Menou et al. (\cite{menou1}) were able to avoid reflares by using a combination
of evaporation (reducing their number) and low quiescent viscosity (increasing
the density at which a heating front is ignited $\Sigma_{\rm max}$). Although
the recurrence timescale predicted between outbursts was reasonable, the
outburst shape did not compare well with FRED-type observed lightcurves, the
rise time being too long.  In addition, the value for the quiescent viscosity
was lower than usually assumed ($\alpha_{\rm c}\sim 5 \times 10^{-3}$). This is
not a problem {\em per se} considering the present knowledge on viscosity in
quiescence (see e.g. Gammie \& Menou 1998) but it could indicate 
different transport mechanisms in
U Gem type dwarf novae with $\alpha_{\rm c}\sim 0.02$ and SXTs with
$\alpha_{\rm c}\sim 0.005$. However, the mechanisms might be the
same, but the efficiencies different. However,
very low values of $\alpha_{\rm c}$ have been advocated in the dwarf nova
WZ Sge (Smak \cite{s93}) which is in many respects similar to SXTs; in
this case, the very low value of $\alpha_{\rm c}$ must be due to a different
mechanism, since it is very unlikely that the efficiency of the same mechanism
could vary by 2 or 3 orders of magnitude. The following section
will show how irradiation by lowering $\Sigma$ behind the cooling front solves
this problem without the need for very low values of $\alpha_{\rm c}$.

\section{Irradiated discs}

\begin{figure} 
%makefig4 sur apollo. Utilise sub142
\resizebox{\hsize}{!}{\includegraphics{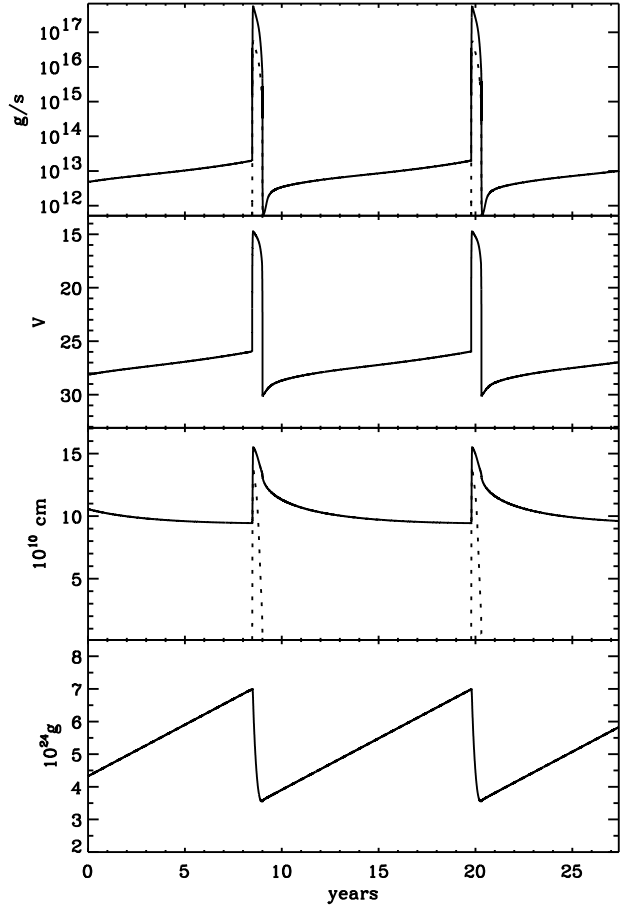}} 
\caption{Example of an outburst cycle when irradiation is
included. From top to bottom: the mass accretion rate at the inner
edge $\dot{M}_{\rm in}$ (full line) and $\dot{M}_{\rm
irr}=\epsilon\dot{M}_{\rm in}$ (dotted line), the V magnitude, the
outer disc radius $R_{\rm out}$ (full line) and the transition radius
$R_{\rm trans}$ between the hot and the cold regions (dotted line),
the mass of the disc $M_{\rm disc}$.  The parameters are
$M_1=7$~\msol, $\dot{M}_{\rm tr}=10^{16}{\rm ~g}\;{\rm s}^{-1}$,
$\alpha_{\rm h}=0.2$, $\alpha_{\rm c}=0.02$, $<R_{\rm
out}>=10^{11}$~cm and a fixed inner radius at $R_{\rm
in}=10^9$~cm. Details of the outbursts and the evolution of the
density and temperature profiles can be seen in
Figs.~\ref{fig:riseidim}-\ref{fig:quietidim}.}
\label{fig:idim} 
\end{figure}

In this section we study the effect of including self-irradiation to
the standard DIM. We present only one model for this purpose (others
may be found in Dubus \cite{dubus0}) with $M_1=7$~\msol, $\dot{M}_{\rm
tr}=10^{16}{\rm ~g}\;{\rm s}^{-1}$, $\alpha_{\rm h}=0.2$, $\alpha_{\rm
c}=0.02$, $<R_{\rm out}>=10^{11}$~cm and a fixed inner radius at
$R_{\rm in}=10^9$~cm (this is further discussed below in
\S\ref{p:quiet}).  Irradiation is included using Eq.~\ref{ill} with
$\epsilon=0.1$ when $\dot{M}_{\rm in}>10^{16}$~g$\;$s$^{-1}$ and
$\epsilon=0.1 (\dot{M}_{\rm in}/10^{16}{\rm ~g~s}^{-1})^6$ below.
This amounts to a quick cutoff of irradiation below $10^{16}{\rm
~g~s}^{-1}$. The exact form of $\epsilon$ matters little as long as it
ensures that irradiation is only important during the outburst (a
reasonable assumption). The resulting outburst time profile is shown
in Fig.~\ref{fig:idim}.

\subsection{Rise\label{p:rise}}

We start with a cold, quiescent disc in which most of the matter has
been accreted in a preceding outburst. During quiescence (see below),
mass transfer from the secondary replenishes the flow until the
density becomes high enough and an annulus arrives at
the thermally unstable branch of the \bS-curve. The ring cannot
maintain thermal equilibrium and undergoes a transition to the
hot branch.

As in the standard DIM, two heating fronts are formed which propagate
inwards and outwards from the ignition radius (Menou et
al. \cite{menou2} and references therein).  In `inside-out' (type B)
outbursts the ignition radius is small and the propagation time of the
inward bound front is very small compared to the other one. In
`outside-in' (type A) outbursts, the opposite is true. In the standard
DIM, `inside-out' fronts can stall rather easily leading to short,
low-amplitude outbursts. The density profile in quiescence is
generally not far from $\Sigma_{\rm max}\propto R$ so that, as the
outward front progresses through the disc, it encounters regions of
higher densities and $\Sigma_{\rm max}$ (the critical density needed
to raise a ring to the hot branch). If the front does not transport
enough matter to raise the density at some radius above $\Sigma_{\rm
max}$, it stalls and a cooling front develops. `Outside-in' fronts
always progress through regions of decreasing $\Sigma_{\rm max}$ and
therefore always heat the whole disc. Another consequence is that
inside-out fronts propagate slowly, leading to slow rise times. This
is particularly evident in Fig.~8 of Menou et al. (\cite{menou1}).

\begin{figure}
% makefig5a sur apollo.
\resizebox{\hsize}{!}{\includegraphics{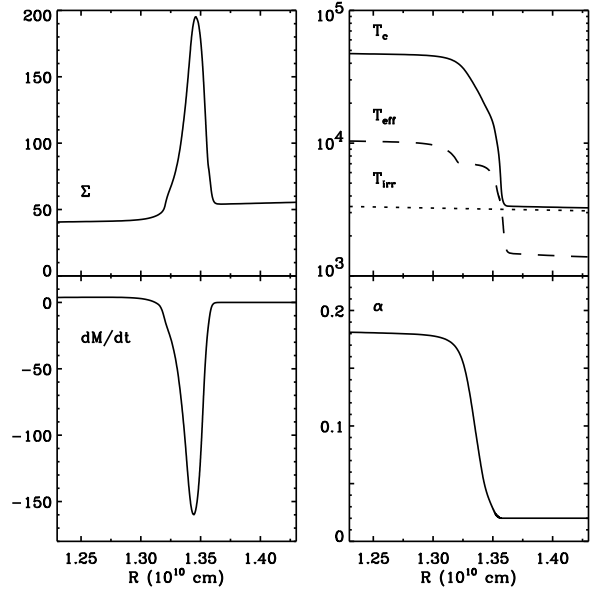}}
\caption{An `inside-out' heating front in an irradiated disc. This is
the front at $t\approx1.5$~days in Fig.~\ref{fig:riseidim}. $\Sigma$
is in g$\;$cm$^{-2}$, temperatures are in K, $\dot{M}$ is in units
of $10^{16}$~g$\;$s$^{-1}$.  The cold outer disc is almost
isothermal with $T_{\rm c}\approx T_{\rm irr}$.}
\label{heating} 
\end{figure}

\begin{figure*}[h]
	\resizebox{11cm}{!}{\includegraphics{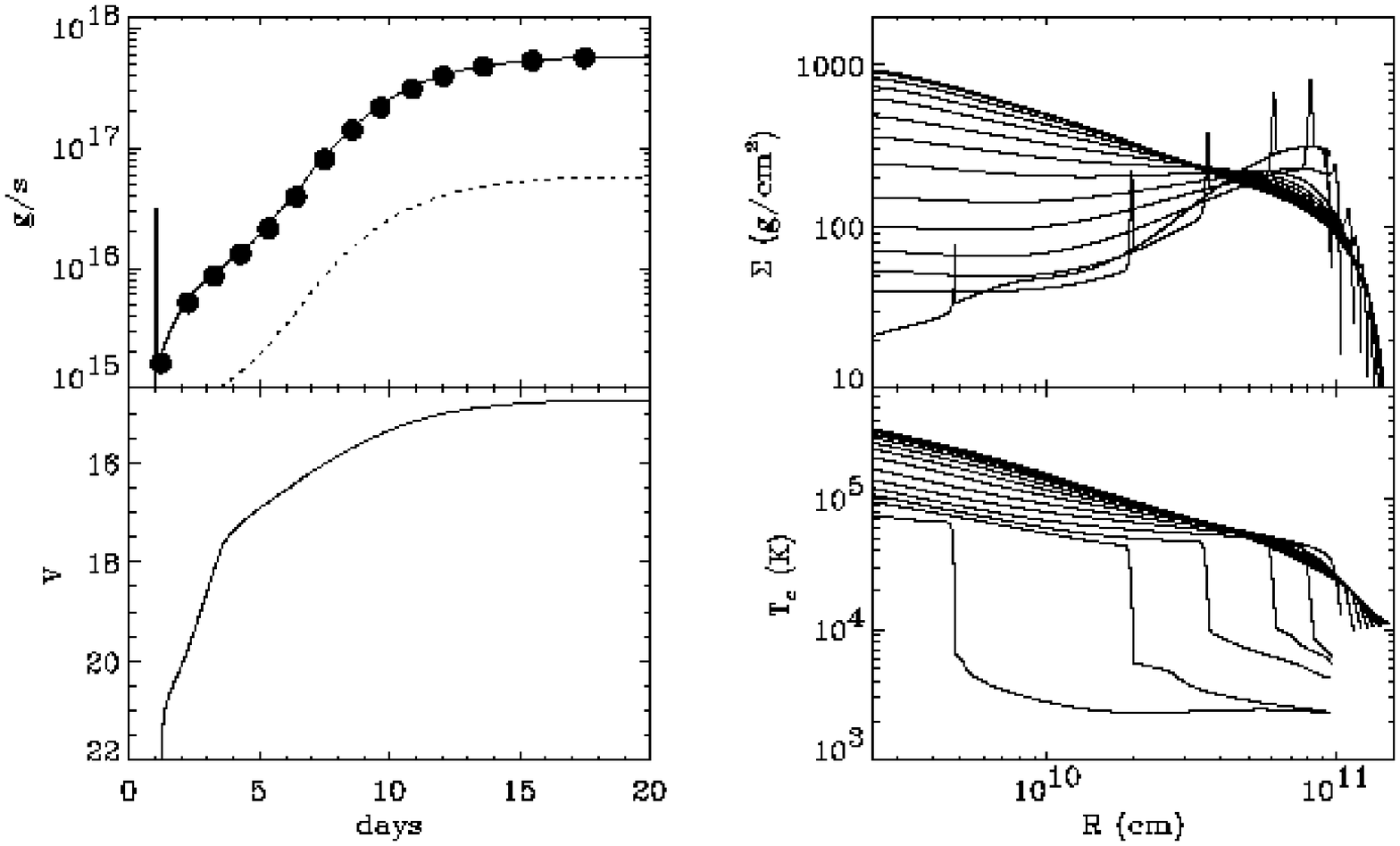}} \hfill
	\parbox[b]{55mm}{ \caption{The outburst rise for the model of
	\S4 (irradiated, with a fixed $R_{\rm in}$). The upper left
	panel shows $\dot{M}_{\rm in}$ and $\dot{M}_{\rm irr}$ (dotted
	line); the bottom left panel shows the $V$ magnitude. The
	spike at $t\approx 1$~day corresponds to the arrival at
	$R_{\rm in}$ of the inward propagating front. The rise is
	dominated by the outward front (inside-out outburst). Each dot
	corresponds to one of the $\Sigma$ and $T_{\rm c}$ profiles in
	the right panels. The disc expands during the outburst to
	transport the angular momentum of the material being
	accreted. The profiles close to the peak are those of a
	steady-state disc ($\Sigma\propto T_{\rm c}\propto
	R^{-3/4}$).}
\label{fig:riseidim}}
\end{figure*}

Irradiation does not change the structure of the heating front
(compare Fig.~\ref{heating} with Fig.~1 of Menou et al.
\cite{menou2}) but does change the maximum radius to which an inside-out
outburst can propagate. Since we do not consider self-screening, the
outer disc always sees the irradiation flux from the central
regions. As an inside-out front propagates, $\dot{M}_{\rm in}$ rises
and the outer cold disc is increasingly irradiated (see
Fig.~\ref{fig:riseidim}).  Irradiation heating reduces the critical
density needed to reach the hot branch (Eq.~\ref{simax}), easing front
propagation. Obviously, a larger hot region implies a greater optical
flux and irradiation always lowers the peak optical magnitude.

In an inside-out outburst, the outer regions the front has not yet
reached are frozen on the timescale on which $\dot{M}_{\rm in}$
evolves (see Fig.~\ref{fig:riseidim}). If $\dot{M}_{\rm in}$ (hence
$\dot{M}_{\rm irr}$) increases on timescales shorter than the thermal
timescale in the cold disc this can lead to situations where
$T_{\rm irr}>T_{\rm c}$ at some radii. The vertical heat flux changes
sign and would require negative values of $\alpha$ in our treatment of
thermal imbalance i.e. the assumptions of the code break down. We
assumed that such an annulus is dominated by irradiation i.e. is
isothermal at $T_{\rm c}$ and that irradiation contributes an
additional heating term to the radial thermal equation
(Eq.~\ref{eq:heat}) $Q^+_{\rm add}=\sigma (T_{\rm irr}^4-T_{\rm c}^4)$
to reflect the imbalance at the photosphere between the outgoing flux
$\sigma T^4_{\rm c}$ and the incoming flux $\sigma T^4_{\rm
irr}$. This, or other assumptions, actually had very little influence on
the rise-to-outburst lightcurve. Further studies of this phenomenon
would require a detailed model of irradiation in SXTs where the
geometry, the exact flux and spectrum of irradiating photons are
properly set out. Note also that the absence of any back loop in the
equations to prevent this situation may suggest that some additional
physics is needed. One possibility is that rapidly increasing
irradiation would evaporate the upper layers of the disc (Begelman
et al. \cite{bmks}; Hoshi \cite{hoshi}; Idan \& Shaviv
\cite{idan}; de Kool \& Wickramasinghe \cite{kw}).

Fig.~\ref{fig:riseidim} shows the evolution of the $\Sigma$ and
$T_{\rm c}$ radial profiles during the outburst rise. As the front
reaches the outer edge of the disc, the profiles in the hot region
converge to those of a steady disc with constant $\dot{M}(R)$. This is
because the viscous timescale, which is inversely proportional to
$\alpha T_{\rm c}$, becomes short enough to equilibrate the mass flow
in the hot region (e.g. Menou et al. \cite{menou2}, and
references therein).  Irradiation has little influence on the actual
vertical structure in this region as discussed in Dubus et
al. (\cite{dubus}) and we find $T_{\rm c}\propto\Sigma\propto
R^{-3/4}$ as in a non-irradiated steady disc. Only in the outermost
disc regions does the vertical structure become
irradiation-dominated, i.e. isothermal.

The peak accretion rate (and optical magnitudes) is rather difficult
to estimate analytically since it depends on the maximum radius to
which the heat front can propagate. Qualitatively, this is set mostly
by the conditions in the disc at the onset of the outburst. The
principle factor is the total disc mass when the outburst starts
$M_{\rm max}$ which is constrained by $\Sigma_{\rm max}$.  Parameter
studies (\S6) show that any changes which result in a lower
$\Sigma_{\rm max}$, hence in a lower $M_{\rm disc}$ do lead to smaller
outburst peaks. Increasing $M_1$, $\alpha_{\rm c}$
\footnote{The dependence is really on the ratio $\alpha_{\rm
h}/\alpha_{\rm c}$. The mass in the disc is almost the same between
the beginning and the peak of the outburst. Assuming $M_{\rm
max}\propto \Sigma_{\rm max}\propto \alpha_{\rm c}^{-0.8}$ and $M_{\rm
max}\propto \dot{M}_{\rm peak}^{0.7}\alpha_{\rm h}^{-0.8}$
(Shakura-Sunyaev steady disc) shows $\dot{M}_{\rm peak}\propto
(\alpha_{\rm h}/\alpha_{\rm c})^{8/7}$.}  or decreasing $R_{\rm out}$
give smaller $\dot{M}_{\rm peak}$. Increasing the mass transfer rate
leads to higher disc masses in quiescence and higher peaks.

In any case, $\dot{M}_{\rm peak}$ should be lower than or of order of
the critical mass accretion rate $\dot{M}_{\rm crit}$ corresponding to
$\Sigma_{\rm min}$ at $R_{\rm out}$, as the disc is close to steady
state at maximum if it has been brought entirely to a hot state. If
the heating front has not been able to reach the outer edge,
$\dot{M}_{\rm peak}$ will be lower.

\subsection{Decay from outburst maximum} 

The outburst decay can generally be divided into three parts:
\begin{itemize}
\item First, outer disc X-ray irradiation inhibits the cooling-front
propagation.  Since more mass is lost than gained from the secondary
($\dot{M}_{\rm in}\gg\dot{M}_{\rm tr}$) the disc is drained by viscous
accretion of matter (King \& Ritter 1998).

\item Second, the accretion rate becomes low enough that the
X-ray irradiation is unable to prevent the cooling from
propagating. The propagation speed of this front is, however,
controlled by irradiation.

\item Irradiation plays no role and the cooling front
switches off the outburst on a local thermal time.
\end{itemize}

We now discuss these three phases in more detail.

\subsubsection{Viscous ``exponential'' decay\label{p:viscous}}

In the model presented here the disc becomes fully ionized at the
outburst peak. The disc then evolves with $\dot{M}(R)$ almost constant
so that $\nu \Sigma \sim \dot{M}_{\rm in}(t)/3\pi$; the total mass in
the disc is thus
\begin{equation}
M_{\rm d}=\int 2\pi R\Sigma dR \propto \dot{M}_{\rm in} \int {2\over 3} {R
\over \nu} dR
\label{eq:expop}
\end{equation}
and one can consider that the disc decays through a sequence of
quasi-stationary states.  In irradiation dominated discs as well as in
standard Shakura-Sunyaev discs, $\nu \propto T \propto
\dot{M}^{\beta/\left(1+\beta\right)}$, with $\beta = 3/7$ in hot
Shakura-Sunyaev discs, and $\beta$ = 1/3 in irradiation dominated
discs. The outer disc radius does not vary much so the time evolution
of the disc mass is:
\begin{equation}
\frac{{\rm d} M_{\rm d}}{{\rm d} t}=-\dot{M}_{\rm in}\propto M_{\rm d}^{1+\beta},
\label{eq:expo}
\end{equation}
showing that $\dot{M}_{\rm in}$ evolves almost exponentially, as long
as $\dot{M}_{\rm in}^\beta$ can be considered as constant (i.e. over
about a decade in $\dot{M}_{\rm in}$, as found by King \& Ritter
\cite{kr} for irradiated discs, and by Mineshige et al. \cite{myi}
for non-irradiated discs with angular momentum removal). Power-laws
with indices close to -1 are found in discs with constant angular
momentum, i.e. in which there are no tidal torques preventing the
outer disc radius to expand indefinitely (Lyubarskii \& Shakura
\cite{ls}; Cannizzo et al. \cite{clg}; Mineshige et al.
\cite{myi}), and are of no practical interest here.

The decay is viscous as long as thermal equilibrium can be maintained.
Incidentally, this makes it difficult, if not impossible, to have a
viscous decay in the standard DIM without additional assumptions. In
outside-in outbursts, matter accumulates at the outer disc edge and the
resulting density excess decreases viscously, producing `flat-top' light
curves. Thermal equilibrium requires the whole disc to be kept in the
hot state. If the outer disc is cold then there will be a thermally
unstable region evolving rapidly (i.e. non-viscously, see next
paragraph). Keeping a disc hot can be achieved by additional mass
transfer during the outburst or non-standard sources of heating (e.g.
tidal heating as in Buat-M\'enard et al. \cite{bhl1}).
A smaller $\alpha_{\rm c}$ increases the difference between 
$\Sigma_{\rm max}$ and $\Sigma_{\rm min}$ and hence can also help keep
the disc in the hot state longer (Menou et al. 2000). 

Irradiation heating not only leads to the exponential decay but also
provides a natural way to keep the disc hot. An irradiation
temperature at the outer edge above $10^4$~K ensures hydrogen is
everywhere ionized. For such temperatures, the disc is thermally
stable whatever the local density $\Sigma$ and accretion rate. Put
differently, the lower branch of the \bS-curve disappears when $T_{\rm
irr}\gta10^4$~K (see Fig.~4-5 of Dubus et al. \cite{dubus}) or
$\Sigma_{\rm min}=\Sigma_{\rm max}$ at $T_{\rm irr}\approx10^4$~K (and
are undefined below: see Eq.~\ref{simax},~\ref{simin}).

One should remember, however, that many SXT light-curves are {\em not}
exponential despite their discs being irradiated. Clearly additional
physical processes have to be taken into account if one wishes to explain
such `non-typical' behaviour (see e.g. Esin et al. \cite{esin}).

\begin{figure*}[h]
	\resizebox{11cm}{!}{\includegraphics{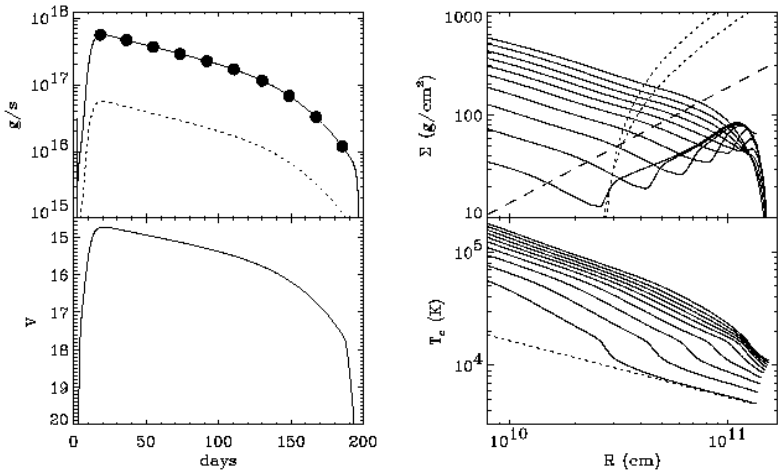}} \hfill
	\parbox[b]{55mm}{ \caption{The outburst decay for the model
	discussed in \S\ref{p:viscous}-\ref{p:linear}. The fully
	ionized disc decays viscously (exponential lightcurve) until
	$t\approx70$~d where $T_{\rm irr}(R_{\rm out})\approx
	10^4$~K. Thereafter the outer disc can fall on the cold branch
	of the \bS-curve and a cooling front appears, propagating
	inward only as far in as $T_{\rm irr}\approx 10^4$~K
	(irradiation-controlled linear decay). At $t\approx190$~d
	irradiation shuts off and the disc cools quickly. The dashed
	line is $\Sigma_{\rm min}$ for $\alpha_h$ without irradiation,
	showing the post-front $\Sigma$ are very low when irradiation
	is included. $\Sigma_{\rm min, max}(\alpha,T_{\rm irr})$, and
	$T_{\rm irr}$ are shown for the last $\Sigma$ and $T_{\rm c}$
	profiles (dotted lines). }
\label{fig:decayidim}}
\end{figure*}
\begin{figure*}[h]
	\resizebox{11cm}{!}{\includegraphics{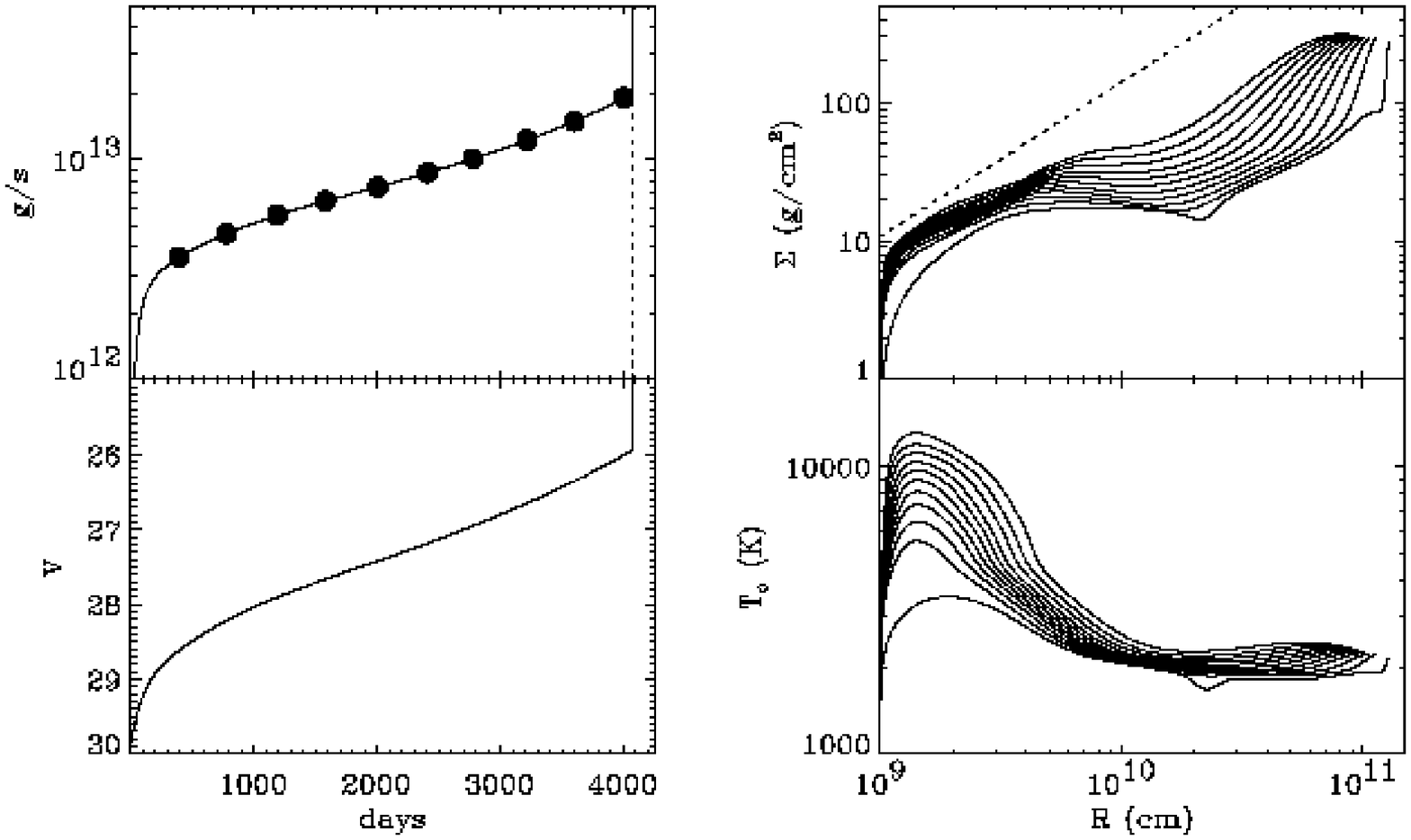}} \hfill
	\parbox[b]{55mm}{ \caption{Quiescence for the model discussed
	in \S\ref{p:quiet}. The irradiation flux is negligible in
	quiescence and the overall evolution is the same as in the
	standard DIM. Mass transfer from the secondary is slow enough
	that matter diffuses down the disc, gradually increasing
	$\dot{M}_{\rm in}$. The outburst is triggered inside at
	$R\approx 2\times 10^9$~cm when $\Sigma$ reaches $\Sigma_{\rm
	max}$ (dotted line). Lower densities at the beginning of the
	quiescent state (due to the irradiation-controlled outburst
	decay) lead to a long recurrence time (about 10 years).}
	\label{fig:quietidim}}
\end{figure*}

\subsection{Irradiation-controlled linear decay\label{p:linear}}

The second stage of the decay starts when a disc ring cannot maintain
thermal equilibrium and switches to the cool lower branch of the
\bS-curve. The disc outer edge always becomes unstable first with
$\Sigma<\Sigma_{\rm min}$ since $\Sigma\propto R^{-3/4}$ (see above)
while $\Sigma_{\rm min}\propto R$. In an irradiated disc this happens
when the central object does not produce enough X-ray flux to keep the
$T_{\rm irr}(R_{\rm out})$ above $10^4$~K so that the low state is
again possible. The material in the low state has a lower viscosity
and piles up, leading to the appearance of a cooling front with a width
$w\propto H$ (Papaloizou \& Pringle \cite{pappri}; Meyer \cite{meyer};
Fig.~4 of Menou et al. \cite{menou2}). The front propagates through
the disc at a speed $V_{\rm front}\approx \alpha_{\rm h} c_s$ (Meyer
1984 ; Vishniac \& Wheeler \cite{vw}). In the standard DIM, the sound
speed depends on the temperature at the transition between the hot and
cold regions and is almost constant as verified by numerical
calculations (Menou et al. \cite{menou2}).

In an irradiated disc, however, the transition between the hot and cold
regions is set by $T_{\rm irr}$ since the cold branch only exists for
$T_{\rm irr}\lta 10^4$~K. Unlike a non-irradiated disc, the cooling
front can propagate only as far as the radius at which $T_{\rm
irr}\approx 10^4$~K, i.e. as far as there is a cold branch to fall onto.
Here also irradiation controls the decay. The hot region stays close to
steady-state but with a shrinking size $R_{\rm hot}\sim \dot{M}_{\rm
in}^{1/2}$ (Eq.~\ref{ill} with $T_{\rm irr}(R_{\rm hot})=
\mathrm{const}$). The mass redistribution during front propagation is
complex with the hot region losing mass both through inflow and outflow
and our models show steeper declines than what some analytic
approximations predict (e.g. King 1998).

\begin{figure}
% makefig5b sur apollo.
\resizebox{\hsize}{!}{\includegraphics{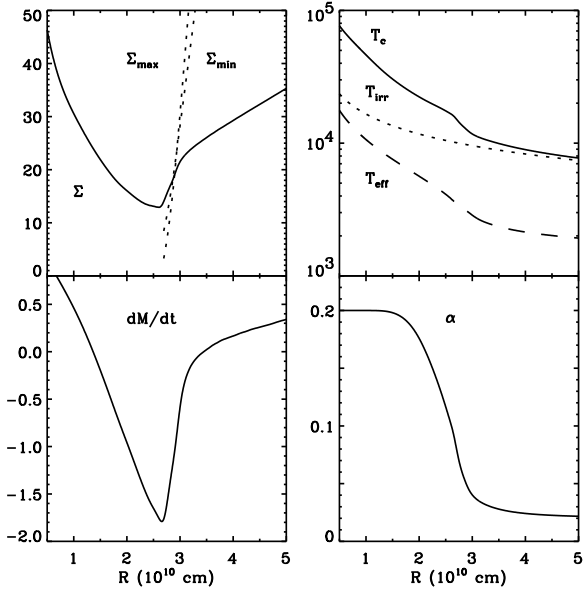}}
\caption{The cooling front in an irradiated disc. This is the front at
$t\approx190$~days in Fig.~\ref{fig:decayidim}. $\Sigma$ is in
g$\;$cm$^{-2}$, temperatures are in K, $\dot{M}$ is in units of
$10^{16}$~g$\;$s$^{-1}$. The front is at the position for which
$T_{\rm irr}$ (dotted line) is $\approx 10^4$~K. At this point
$\Sigma_{\rm min}\approx\Sigma_{\rm max}$ since there is no cold
branch for higher $T_{\rm irr}$ (smaller $R$). As $T_{\rm irr}$
decreases the two critical $\Sigma$ separate and converge to their
non-irradiated values (this can be seen in Fig.~\ref{fig:decayidim}; 
note the values of $\Sigma_{\rm max}$ and $\Sigma_{\rm min}$ shown here
are only accurate to the precision of the fits given by
Eq.~\ref{simax},\ref{simin}).  The cold outer disc is almost
isothermal with $T_{\rm c}\approx T_{\rm irr}$.  The negative values
of $\dot{M}$ show the cooling front transports matter to the outer
disc.}
\label{cooling} 
\end{figure}

The linear decay is illustrated in Fig.~\ref{fig:decayidim}.  The
irradiation temperature is lower than $10^4$~K at the outer edge
for $t>70$~days where the decay becomes steeper. The cooling front (see
also Fig.~\ref{cooling}) appears as a depression propagating inwards
in the surface density profile. The irradiation temperature and the
critical densities $\Sigma_{\rm min,max}$ are plotted for the last
profile at $t\approx 190$~days, showing the cooling front is at the
radius for which $T_{\rm irr} \approx 10^4$~K and that $\Sigma_{\rm
min,max}$ are undefined for lower radii. 

\subsection{Final thermal decay} 

The quick decay after
$t>190$~days is due to vanishing irradiation as $\epsilon$ becomes
very small for $\dot{M}_{\rm in}<10^{16}$~g$\;$s$^{-1}$. 
The cooling front thereafter propagates freely inwards, on a thermal time
scale.

\subsection{Mini-reflares\label{p:mini}}

As discussed above, the edge of the hot zone is at the stability limit
set by $T_{\rm irr}\approx 10^{4}$~K. At this point, one has
$\Sigma\approx\Sigma_{\rm min}\approx\Sigma_{\rm max}$ because
irradiation modifies the critical surface densities.  Further away
from the edge of the hot zone, the irradiation temperature decreases
rapidly ($T^4_{\rm irr}\propto R^{-2}$) and $\Sigma_{\rm max}$
gradually becomes greater than $\Sigma_{\rm min}$. In the
non-irradiated limit, the ratio $\Sigma_{\rm max}/\Sigma_{\rm min}$ is
a constant depending only on the ratio $\alpha_{\rm c}/\alpha_{\rm h}$
(see e.g. Fig.~\ref{fig:reflare}).

The region immediately behind the cooling front where $\Sigma_{\rm
max}$ is close to $\Sigma_{\rm min}$ is clearly very unstable. Slight
variations of $\Sigma$ may suffice to have $\Sigma>\Sigma_{\rm max}$
behind the cooling front and therefore start a reflare just as in
a non-irradiated disc. This depends on the numerical details
of the model, such as the functional of $\alpha$ with temperature 
(\S\ref{talpha}).

\begin{figure}
% makemini sur apollo.
%\resizebox{\hsize}{!}{\includegraphics{miniref.eps}}
\includegraphics{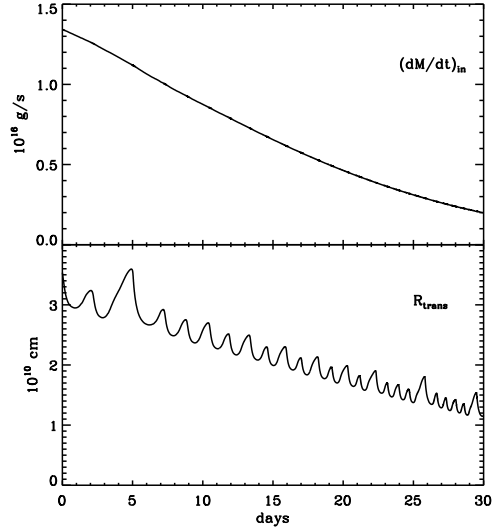}
\caption{Example of mini-reflares in an irradiated disc. The region
close to $T_{\rm irr}\approx 10^4$~K is highly unstable with
$\Sigma_{\rm min}\approx\Sigma_{\rm max}$ (Fig.~\ref{cooling}). This
can lead to a succession of heat fronts propagating quickly back and
forth in a small region (bottom panel). The exact details depend on
the model assumptions (e.g. $\alpha$). This has no influence on
either $\dot{M}_{\rm in}$ (top panel) or the optical flux (not shown).
See \S\ref{p:mini} for details.}
\label{miniref} 
\end{figure}

These {\em mini-reflares} are unimportant. The critical density
$\Sigma_{\rm min}$ increases quickly behind the cooling front (see
Fig.~\ref{cooling}) because the irradiation flux decreases with
$R^{-2}$. Therefore, the heating front in a mini-reflare reaches almost
immediately a radius where $\Sigma<\Sigma_{\rm min}$ and cooling
resumes. Furthermore, irradiation seriously depletes the disc and the
front finds very little matter to fuel its propagation. $\Sigma_{\rm
min}$ is much lower than its non-irradiated value
(Fig.~\ref{fig:decayidim}). For instance, Fig.~4 of Dubus et al.
(\cite{dubus}) shows $\Sigma_{\rm min}(T_{\rm irr}\approx
10^4\mathrm{K})\approx30$~g$\;\mathrm{cm}^{-2}$ instead of
$\Sigma_{\rm min}(T_{\rm
irr}=0\mathrm{K})\approx150$~g$\;\mathrm{cm}^{-2}$. The much lower
post-cooling front densities prevent large reflares as in the standard
DIM. In practice, mini-reflares in irradiated discs have no influence on
the lightcurve (very low amplitudes and cycles, see Fig.~\ref{miniref})
but can be a numerical nuisance. 

\subsection{Quiescence\label{p:quiet}}
As discussed above, irradiation dramatically slows down the cooling
process.  The decay lasts much longer and, accordingly, the
total mass accreted during an outburst is much larger.  The densities
after propagation of the cooling front are low so the total mass of
the disc as it enters quiescence is much less than in the
non-irradiated case. 

The low quiescent surface densities imply low midplane temperatures.
The ($\Sigma$,$T_{\rm c}$) \bS-curves show that for low $\Sigma$ and no
irradiation (as appropriate once the disc enters quiescence) the lower
branch is flat with $T_{\rm c}\approx 2000$~K whatever $R$ or
$M_1$ (see e.g. in Figs.~4-5 of Dubus et al. 1999). In contrast, 
the standard DIM leads to higher $\Sigma$, close
to the non-irradiated $\Sigma_{\rm min}$ and thus to $T_{\rm c}\approx
4000$~K (e.g. Osaki \cite{osaki96}, Menou et al. \cite{menou1}).

The critical density $\Sigma_{\rm max}$ that the disc must reach
before an outburst can start will be the same as in the standard DIM
if irradiation in quiescence is negligible.  This is in general true
since $\dot{M}_{\rm in} \propto T^4_{\rm irr}$ is very
small\footnote{The quiescent disc can be irradiation-dominated if
$\cal C$ is very large (see \S\ref{p:varc}).  In this case, the
critical $\Sigma_{\rm max}$ is lowered by irradiation and the
recurrence timescale is shortened.}. Since the quiescent disc after an
irradiation-controlled decay is always less massive, the time to build
up the critical density will be longer.

In the model presented in this section, $T_{\rm irr}=0$ in
quiescence. The evolution of the $\Sigma$ and $T_{\rm c}$ profiles
(Fig.~\ref{fig:quietidim}) shows matter arriving from the secondary
at the outer edge diffuses down the cold disc. The disc contracts due
to addition of this lower angular-momentum material. The temperature is
almost constant in most of the disc with $T_{\rm c}\approx 2000$~K. This
is not true in the later stages of quiescence where $T_{\rm c}$ rises
in the inner disc due to the build up of matter there: the inner rings
move up along the lower ranch of the ($\Sigma$,$T_{\rm c}$) \bS-curve
and out of the flat $T_{\rm c}\approx 2000$~K region discussed above. In
the last profile, $\Sigma$ is close to $\Sigma_{\rm max}$ (dotted line)
and the outburst is ignited at $R\approx 2\times 10^9$~cm.

The ignition radius depends essentially on the mass transfer rate from
the secondary and {the disc's size} {(Smak \cite{smak84}). For low
$\dot{M}_{\rm tr}$, matter drifts down the cold disc and $\Sigma_{\rm
max}$ is reached in the inner disc (inside-out, type B outburst). As
$\dot{M}_{\rm tr}$ increases, the mass accumulation time at the outer
radius can become lower than the drift time, triggering an outburst in
the outer disc (an outside-in, type A outburst). The accumulation time
is (Ichikawa \& Osaki \cite{ichikawa}; see also Osaki
\cite{osaki95,osaki96} and Lasota \cite{review}):
\begin{equation}
t_{\rm acc}={2\pi R_{\rm out}\Sigma_{\rm max}(R_{\rm out}) \over
\dot{M}_{\rm tr}}\Delta R
\end{equation}
A ring of material spreads viscously over $\Delta R\approx (\nu t_{\rm
acc})^{1/2}$ during $t_{\rm acc}$. Setting $t_{\rm acc}=t_{\rm drift}$
gives an estimate of the accretion rate for which the transition from
type B to A outbursts is expected. Using Eq.~\ref{simax} and $t_{\rm
drift}=\delta t_{\rm vis}$ ($\delta$ is a numerical correction factor
introduced by Osaki \cite{osaki96}) outside-in outbursts are therefore
expected when
\begin{eqnarray}
\dot{M}_{\rm tr} & \gta & 2\pi R_{\rm out}\delta^{-1/2} \nu \Sigma_{\rm
max}(R_{\rm out}) \nonumber\\ 
& \gta & 3.3\times 10^{16} \left({\alpha \over 0.02}\right)^{0.2}
\left({R_{\rm out} \over 10^{11}\mathrm{cm}}\right)^{2.6}\nonumber\\
& & \mathrm{~~~~~~~~~~~~~~}\times \left({M_1\over 7 \mathrm{M}_\odot}\right)^{-0.9} 
\left({T_{\rm c} \over 2000\mathrm{K}}\right) \mathrm{\ g\  s}^{-1}
\end{eqnarray}
with $\delta=1$. This is smaller than
the mass accretion rate needed to stabilize an irradiated disc with
these parameters ($\dot{M}_{\rm tr}\gta 10^{17}$~g~s$^{-1}$, Eq.~30
of Dubus et al. \cite{dubus}).  In principle type A outbursts are thus
possible in SXTs, but numerical calculations show that real outside-in
outbursts starting far out in a large disc are difficult to obtain.
Osaki (\cite{osaki96}) finds $\delta\approx 0.05$ provides an adequate
fit to his calculations. The quiescence time of our model (about 4000 days)
implies $t_{\rm drift}\approx 0.1 t_{\rm vis}(R_{\rm out})$ so that 
the required $\dot{M}_{\rm tr}$ for a type A outburst is above the 
accretion rate for which the disc is stable. 

The inner edge of the accretion disc plays a crucial role in
quiescence. For low accretion rates $\Sigma_{\rm max}$ is reached in
the inner disc. Since $\Sigma_{\rm max}\propto R$ the amount of mass
needed to trigger an outburst is reduced when the disc inner radius is
smaller: the higher $R_{\rm in}$ in quiescence, the longer the
recurrence time. Even with $R_{\rm in}=10^9$~cm the model is still
short of providing 
\begin{itemize}

\item{} the long recurrence timescale ($t_{\rm rec}
\approx$ 10~years compared to the tens of years inferred in SXTs) and

\item{} the quiescent X-ray luminosities: the predicted quiescent
accretion rate, in the range $10^{12-13}$~g s$^{-1}$ is still too
small to account for observed X-ray luminosities exceeding $10^{32}$
erg s$^{-1}$.
\end{itemize}
The next section will show that including disc evaporation into a hot,
low-density, accretion flow can solve these problems.

\section{Irradiated discs with evaporation \label{sec:evap}}

Following the work of Menou et al. (\cite{menou1}), we now assume that
the inner disc is gradually evaporated into a hot, radiatively
inefficient accretion flow during the decay from outburst. The exact
nature of the flow is not important as long as it does not participate
in the dynamics of the thin accretion disc. A good assumption is that
the inner disc is replaced by an advection-dominated accretion flow
(ADAF). ADAF+thin disc models have been successful in explaining the
spectral states of SXTs and in particular, the quiescent X-ray flux
provided the mass accretion rate in the ADAF is large enough
(e.g. Esin et al. \cite{esin2}).

The accretion timescale in an ADAF becomes of the order of the thermal
timescale so that the inner flow can be considered as quasi-steady and
does not participate in the limit cycle. A good assumption in our
dynamical study is therefore to treat evaporation as a variation of
the inner radius of the thin disc we model. We take the same rate of
evaporation as in Menou et al. (\cite{menou1}) :
\begin{equation}
\dot M_{\rm ev}(R)=0.08 \dot M_{\rm Edd} \left[ \left( \frac{R}{R_s}
\right)^{1/4} + {\cal E}\left( \frac{R}{800 R_s} \right)^2 \right]^{-1}
\label{evapor}
\end{equation}
The inner radius of the disc is defined as:
\begin{equation}
\dot{M}_{\rm ev}(R_{\rm in})=\dot{M}_{\rm in}
\end{equation}
Below this radius all the matter evaporates. Because of the steep
dependence of $\dot M_{\rm ev}$ on $R$, we can safely neglect
evaporation above this radius. As noted in Menou et
al. (\cite{menou1}), the detailed functional dependence of $\dot
M_{\rm ev}(R)$, or of $R_{\rm in}(\dot M_{\rm in})$ has essentially no
effect on the results; what matters is the value of the inner disc
radius during quiescence.

Equation (\ref{evapor}) is entirely ad hoc, and is not based on
any particular physical mechanism. There are models that, in
principle, allow  for the determination of the evaporation rate
as a function of radius.  However, the variety of models shows that
there is no general agreement on the physical cause of
evaporation. For example, Meyer \& Meyer-Hofmeister (1994) assume that
electron conduction from the disc to the corona plays a major role,
whereas Shaviv, Wickramasinghe \& Wehrse (1999) consider a thermal
instability related to the opacity law. In addition, even for a given
model, the evaporation law depends on parameters which are not easily
measured (as for example the magnetic field and its coherence in the
case of electron conduction ; Meyer, Liu \& Meyer-Hofmeister 2000). As
all these models have been proposed to explain the existence of holes
in observed accretion discs, they, by definition, tend to produce
similar results, i.e. the truncation radius must be at a detectable
distance from the compact objet. And of course, for the same
reason, Eq. (\ref{evapor}) would also reproduce these results. The
latest calculations by Meyer et al. (2000) give evaporation rates of
the order of 1.4 $\times 10^{16}$ g s$^{-1}$ at 10$^{10}$cm from a 6
M$_\odot$ black hole, and 1.6 $\times 10^{17}$ g s$^{-1}$ at 10$^9$
cm. These are larger by a factor $\sim$ 15 and $\sim$ 2 respectively
than the values we are using  here. Note that the
evaporation rate is so high in Meyer et al. (2000) that the 
quiescent disc would  be truncated at too large a radius to be
unstable.

The ADAF efficiency roughly scales as $\dot{M}$ (i.e. the luminosity
scales as $\dot{M}^2$, Esin et al. \cite{esin2}) . From
Eq.\ref{evapor}, this is equivalent to $\epsilon \propto R^{-2}_{\rm
in}$. In the model, we assume $\epsilon=0.1$ when $R_{\rm in}=R_{\rm
min}$ (in outburst) and $\epsilon=0.1\times(R_{\rm min}/R_{\rm
in})^{-2}$ when $R_{\rm in} \ge R_{\rm min}$ (in
quiescence). Irradiation is usually negligible in quiescence so these
assumptions have little importance there. Varying $\epsilon$ may
change slightly the end of the outburst when evaporation becomes
important.

We took $R_{\rm min}=5\times 10^{8}$~cm in all the following models.  As
discussed below in \S\ref{p:outburst}, the only time $R_{\rm min}$ is
reached is in outburst where the inner disc is in thermal equilibrium
and close to steady-state. The inner disc edge plays no role in this
case.

In the following we discuss in some detail the same model as in \S4
but including evaporation. The overall outburst time profile is shown in
Fig.~\ref{fig:tidim}.  A comparison with Fig.~\ref{fig:idim} shows the
outburst is almost identical but that the disc spends much more time
in quiescence ($t_{\rm rec}\approx 21$~years). The maximum luminosity
reached during outburst is also increased.

\begin{figure}
%makefig7 sur apollo. Utilise sub126
\resizebox{\hsize}{!}{\includegraphics{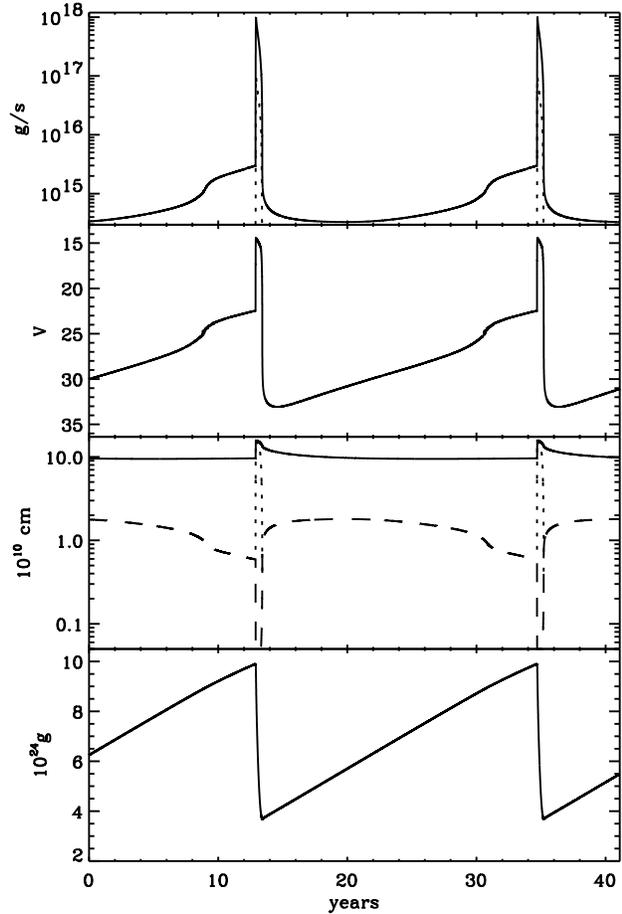}}
\caption{Example of an outburst cycle when irradiation and evaporation
are included. The parameters of the model are the same as in
Fig.~\ref{fig:idim}. From top to bottom: $\dot{M}_{\rm in}$ (full
line) and $\dot{M}_{\rm irr}$ (dotted line); V magnitude; $R_{\rm
out}$ (full line), $R_{\rm trans}$ (dotted line) and $R_{\rm in}$
(dashed line), $M_{\rm disc}$. Details of the outburst and the
evolution of the density and temperature profiles can be seen in
Figs.~\ref{fig:risetidim}-\ref{fig:quiettidim}.}
\label{fig:tidim}
\end{figure}

\subsection{The outburst\label{p:outburst}}

After an outburst is triggered, the mass accretion rate at the inner
edge gradually increases as in the previous model.  The inner radius
of the thin disc decreases until its minimum value is reached for
$\dot{M}_{\rm in}=\dot{M}_{\rm ev}(R_{\rm min})$. A slight change of
slope is associated with this (at the 4th dot in
Fig.~\ref{fig:risetidim}). The time $t_{\rm rise}$ the disc needs to
reach $R_{\rm min}$ is a viscous timescale. Before the onset of the
outburst, $R_{\rm in}\approx 6\times 10^9$~cm. Therefore, $t_{\rm rise}$
is of the order of the difference between $t_{\rm vis}$ at
$6\times 10^9$~cm and $5\times 10^8$~cm. The viscous timescale is given by
\begin{equation} 
t_{\rm vis}= \frac{R^2}{\nu}=(GM_1R)^{1/2}\frac{\mu m_H}{\alpha k T_c}.
\label{eq:tvis}
\end{equation}
For parameters appropriate to this case ($T_{\rm c}\approx 10^5$~K,
$\alpha=0.2$,$\mu=0.5$, $M_1=7$~\msol) the difference gives about 6
days which is in good agreement with the numerical calculation.
Fig.~\ref{fig:risetidim} shows the minimum radius is reached at
$t\approx 5.5$~days, about 3 days after the onset of the outburst.

The subsequent evolution is identical to the model where the inner
radius is kept fixed. The peak accretion rate and optical flux are
higher in the model with evaporation because of the different
quiescent histories of the discs. When the outburst starts, the discs
do not have the same density distribution and total mass
(Fig.~\ref{fig:idim}, \ref{fig:tidim}).

The decay from outburst is also identical to the previous model with a
viscous and a linear decay (Fig.~\ref{fig:decaytidim}). The linear
decay has a limited influence on the lightcurve since the cooling
front only propagates to $8\times 10^{10}$~cm before evaporation sets in
($t\approx 160$~days in Fig.~\ref{fig:decaytidim}). Irradiation then
shuts off with the decreasing $\epsilon$ and the front cools the whole
disc quickly. At the same time, the inner disc radius rises with the
increasing evaporation which also quickens the cooling. In general,
evaporation and/or changes in the irradiation efficiency $\epsilon$
take place for $\dot{M}_{\rm in}\sim\dot{M}_{\rm tr}$ at which point
the total disc mass is close to its minimum value (the disc starts
filling in again) and there is little matter left to accrete.
Therefore, the exact prescriptions for $\dot{M}_{\rm evap}$ and
$\epsilon$ do not have a significant impact on the outburst
lightcurve. For instance, these produce a difference of only a few
days in the outburst lengths between the model of \S4 and the model of
this section.

\subsection{Quiescence with evaporation\label{p:quietevap}}

The whole disc becomes cold and enters quiescence when the cooling
front reaches the varying inner radius $R_{\rm in}$. In
Fig.~\ref{fig:decaytidim} this happens at $t\approx 195$~days when
$R_{\rm in}\approx 5\times 10^{10}$~cm and $\dot{M}_{\rm in}\approx
5\times 10^{15}$~g$\;$s$^{-1}$. The drift time of cold matter is long
and the disc cannot maintain this high accretion rate: $\dot{M}_{\rm
in}$ decreases as material is gradually accreted. This adjustment
happens on a long timescale (cold material) which depends on the
$\Sigma$ profile left after the outburst. With a fixed $R_{\rm in}$,
$\dot{M}_{\rm in}$ is already very low (about
$10^{12}$~g$\;$s$^{-1}$, Fig.~\ref{fig:quietidim}) when the whole
disc becomes cold. This is actually a necessary condition for the disc
at low radii to reach the cold branch. For such low accretion, rates the
supply of material is enough for $\dot{M}_{\rm in}$ to increase steadily
throughout quiescence.

This first stage lasts until material from the outer edge had enough
time to drift to the inner edge and increase $\dot{M}_{\rm in}$. In
Fig.~\ref{fig:quiettidim}\ this happens at $t\approx 3000$~days. The
first three $\Sigma$ radial profiles of Fig.~\ref{fig:quiettidim} show
clearly mass from the outer edge diffusing inwards and reaching the
inner edge at $t\approx 2500$~days. The disc then steadily builds up
mass until the combination of a decreasing $R_{\rm in}$ (hence
$\Sigma_{\rm max}$) and increasing $\Sigma$ triggers an outburst 
(see also Cannizzo \cite{cannizzo3}, Meyer \& Meyer-Hofmeister
\cite{meyer4}).

Under most circumstances (\S\ref{p:quiet}) the outburst will be of the
inside-out type. The reason is that, because of the very large disc
sizes of SXTs, the time it takes for matter to diffuse down the disc
is shorter than the accumulation time at the outer disc.  A truncated
disc does not change this conclusion: evaporation does not prevent
inside-out type B outbursts. 

However, evaporation does suppress the small mini-outbursts which are
found when the inner disc radius is small (see e.g. Hameury et
al. 1998).  With $R_{\rm in}$ fixed at $10^9$~cm, $\Sigma_{\rm max}$
is small, of the order of 10~g$\;$cm$^{-2}$, and an outburst can be
triggered easily as soon as matter diffuses in to the inner
radii. This leads to sequences of mini-outbursts which are not
observed.  A truncated disc will need to build up more mass before an
outburst can start. The critical density $\Sigma_{\rm max}$ varies
with $R$ and will be much higher when $R_{\rm in}$ is large, of the
order of a 100~g$\;$cm$^{-2}$ at $10^{10}$~cm. This prevents small
outbursts from being triggered.

A truncated disc at the onset of an outburst will be more massive than
a non-truncated disc (compare Figs.~\ref{fig:idim} and
\ref{fig:tidim}). But since the disc in outburst cools under the same
conditions (dictated by $T_{\rm irr}$), the mass of the disc at the
end of the outburst will be roughly the same in both cases. If $\Delta
M$ is the mass accreted during the outburst then we have $\Delta
M_{\rm trunc} > \Delta M_{\rm no~trunc}$.

The quiescence time is the time it takes to replenish the mass lost
during the outburst:
\begin{equation}
t_{\rm quiesc}=\frac{\Delta M}{\dot{M}_{\rm tr}-\dot{M}_{\rm in}}
\label{eq:trec}
\end{equation}
Irradiation depletes the disc during the decay, increases
$\Delta M$ and leads to longer $t_{\rm quiesc}$.  In contrast
to models in which the disc extends to low radii, truncated
discs can have larger $\dot{M}_{\rm in}$ in quiescence implying longer
$t_{\rm quiesc}$.  Eq.~\ref{eq:trec} also shows that the
quiescence time depends on the mass transfer rate even for inside-out
outbursts when $\dot{M}_{\rm in}$ in quiescence is a significant
fraction of $\dot{M}_{\rm tr}$. In a disc with low quiescent
$\dot{M}_{\rm in}$, $t_{\rm quiesc}$ for inside-out outbursts is of
the order of $t_{\rm diff}$ which is independent of $\dot{M}_{\rm tr}$
(Osaki \cite{osaki96}; Smak \cite{s93}).

\begin{figure*}[h]
	\resizebox{11cm}{!}{\includegraphics{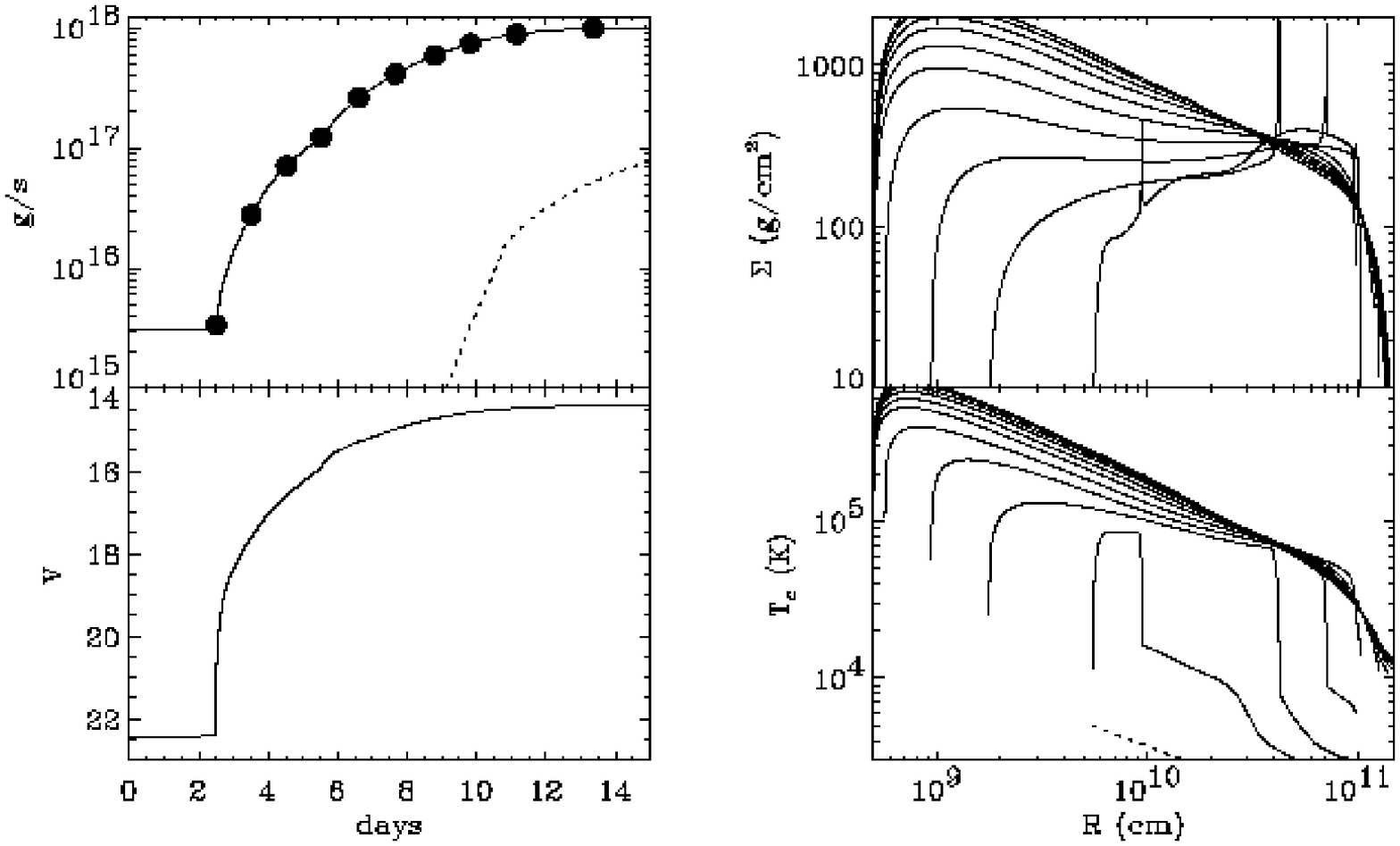}} \hfill
	\parbox[b]{55mm}{\caption{The outburst rise for the model of
	\S5 which includes irradiation and evaporation (see
	\S\ref{p:outburst}). Upper left panel shows $\dot{M}_{\rm in}$
	and $\dot{M}_{\rm irr}$ (dotted line); bottom left panel shows
	the $V$ magnitude. Right panels show radial profiles at
	the different times indicated by the black dots on the
	curves.  Evaporation decreases during the first part of
	the rise with the thin disc extending to smaller radii.  At
	$t\approx 5.5$~days the thin disc reaches the minimum
	possible inner radius of the model. The disc then behaves in
	exactly the same way as in Fig.~\ref{fig:riseidim}.}
	\label{fig:risetidim}}
\end{figure*}
\begin{figure*}[h]
	\resizebox{11cm}{!}{\includegraphics{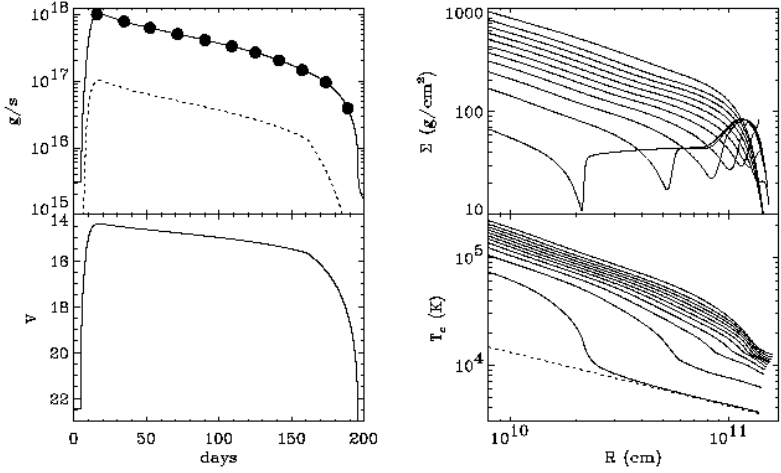}} \hfill
	\parbox[b]{55mm}{\caption{The outburst decay for the model
	discussed in \S\ref{p:outburst}. The disc behaves in the same
	way as in Fig.~\ref{fig:decayidim} until evaporation sets in
	at $t\approx170$~days ($\dot{M}_{\rm in}=\dot{M}_{\rm
	evap}(R_{\rm min})$). This cuts off irradiation and the disc
	cools quickly. In contrast to Fig.~\ref{fig:decayidim}, the
	irradiation cutoff happens before the cooling front could
	propagate through most of the disc, hence the
	irradiation-controlled linear decay ($t\approx 80-170$~days)
	is less obvious in the time profile.  $T_{\rm irr}$ (dotted
	line) is shown for the last temperature profile.}
	\label{fig:decaytidim}}
\end{figure*}
\begin{figure*}[h]
	\resizebox{11cm}{!}{\includegraphics{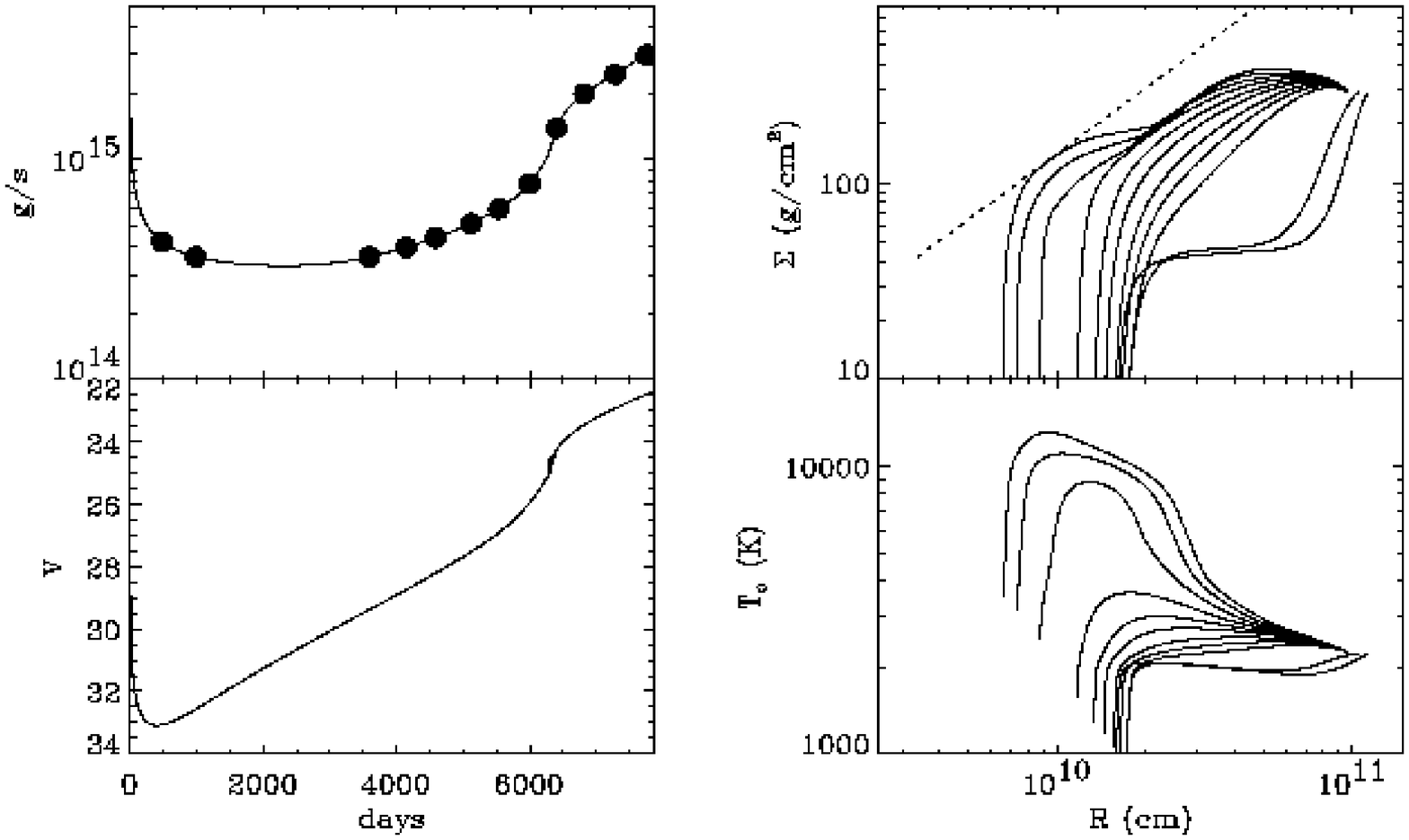}} \hfill
	\parbox[b]{55mm}{\caption{Quiescence for the model discussed
	in \S\ref{p:quietevap}. The irradiation flux is negligible in
	quiescence. Overall, the evolution is the same as in the
	standard DIM. Mass transfer from the secondary is slow enough
	that matter diffuses down the disc, gradually increasing
	$\dot{M}_{\rm in}$ (and lowering $R_{\rm in}$).  The outburst
	is triggered at $R\approx 10^{10}$~cm when $\Sigma$ reaches
	$\Sigma_{\rm max}$ (dotted line). Lower densities at the
	beginning of the quiescent state (due to the
	irradiation-controlled outburst decay) lead to a long
	recurrence time (about 35 years).}  \label{fig:quiettidim}}
\end{figure*}

\section{Parameter study}

\subsection{Changing the irradiation strength $\cal C$\label{p:varc}}

We show in Fig.~\ref{irrcl} the effects of a stronger or weaker
irradiation on the outbursts.  The strength of irradiation is set by
the value of $\cal C$ and we show models for ${\cal C}=10^{-3}{\rm ,\
}5\times10^{-3}{\rm ,\ }10^{-2}{\rm ,\ }10^{-1}$.  As irradiation
grows stronger, $\Sigma_{\rm min}$ becomes lower (see Eq.~\ref{simin})
and the first effect is to increase the length of the outburst: more
mass gets accreted before $\Sigma_{\rm min}$ is reached.  Thus, the
relative amount of matter accreted during outburst, $\Delta M/M$,
increases with $\cal C$.  The disc spends more time accreting in
quiescence before it can build up enough density somewhere to reach
$\Sigma_{\rm max}$.

\begin{figure}
% makefig6 sur apollo
\resizebox{\hsize}{!}{\includegraphics{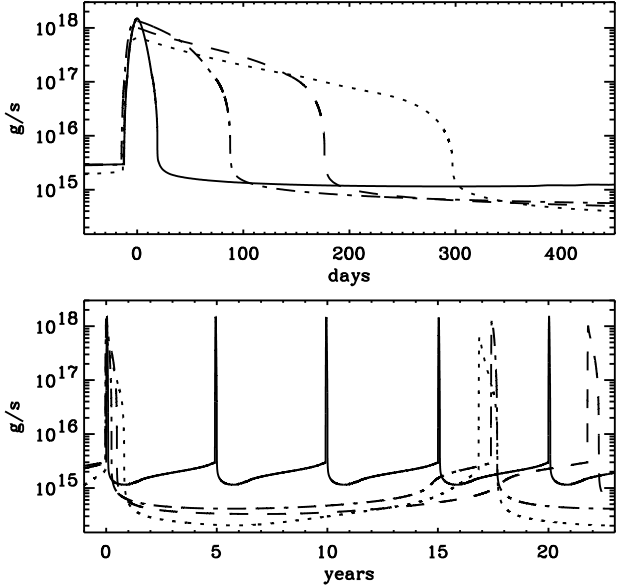}}
\caption{Effect of changing the irradiation parameter $\cal C$ on the
outburst time evolution and cycle (see \S\ref{p:varc} for
details). ${\cal C} / 10^{-3}=$0 (full line), 1 (dash-dotted), 5
(dashed), 100 (dotted). All other parameters are identical to those of
the model described in \S4. The non-irradiated curve shows no reflares
in this particular case. This is in general not the case, even with
evaporation: the same model with $M_1=6$\msol\ shows them.}
\label{irrcl} 
\end{figure} 

Like $\Sigma_{\rm min}$, $\Sigma_{\rm max}$ also decreases with
increasing irradiation (see Eq.\ref{simax}). Therefore, one would also
expect $\Sigma_{\rm max}$ to be reached more easily, hence to have
shorter quiescent times. Usually the irradiation flux is low at the
end of quiescence so only large values of $\cal C$ can heat the disc
enough to change $\Sigma_{\rm max}$ (see Fig.~\ref{irrch}).

There are two competing effects: (i) the increased amount of matter
accreted during an outburst lengthens the quiescence time; (ii)
unusually strong irradiation in quiescence lowers the critical
density needed to reach an outburst, which reduces the quiescence
time. This happens when $\cal C$ is large, in which case irradiation
is no longer negligible in quiescence (in contrast to the models of
\S4-5). The numerical models show that there is a ${\cal C}_{\rm max}$
for which the recurrence time is greatest, ${\cal C}_{\rm max}\approx
5\times10^{-3}$, i.e. the `standard' value for which persistent
low-mass X-ray binaries are stabilized (Dubus et
al. \cite{dubus}). This is probably no more than a coincidence as
${\cal C}_{\rm max}$ certainly changes as a function of $M_1$, $R_{\rm
out}$, etc.

The average mass in the disc $<$$M_{\rm d}$$>$ decreases with stronger
irradiation. $<$$M_{\rm d}$$>$ depends mostly on the mass of the disc
in quiescence. Since $\Sigma<\Sigma_{\rm max}$ in quiescence and
$\Sigma_{\rm max}$ decreases with irradiation, $<$$M_{\rm d}$$>$
decreases as well. Another consequence of stronger irradiation are
the higher optical fluxes in outburst due to the higher outer disc 
temperatures.

\begin{figure}
% makefigc.pro sur hermitage. avec rad126 et rad129
\includegraphics{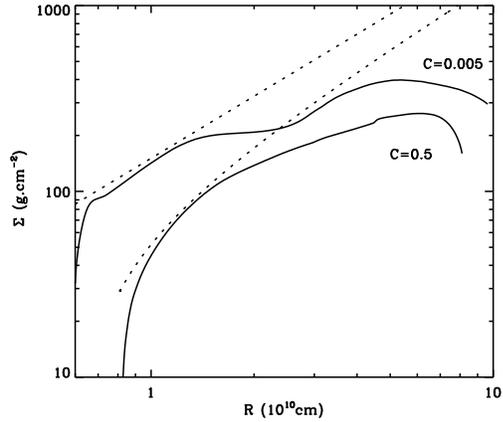}
\caption{Density profiles $\Sigma(R)$ in quiescent evaporated discs
before the onset of an outburst. The dotted line is $\Sigma_{\rm
max}$. For ${\cal C}=5\times10^{-3}$ irradiation is negligible in
quiescence and the critical $\Sigma_{\rm max}$ is the same as for a
non-irradiated disc. For larger values of $\cal C$ irradiation becomes
important, lowering $\Sigma_{\rm max}$. The critical density the disc
must reach before an outburst is significantly less and the recurrence
time between outbursts is decreased. See \S\ref{p:varc} for details.}
\label{irrch} 
\end{figure} 

\subsection{Changing the viscosity $\alpha$ \label{p:vara}}
The effect of changing $\alpha_{\rm h}$ and $\alpha_{\rm c}$ is
straightforward. The $\alpha$ parameter sets the viscous time of the
accretion flow (Eq.~\ref{eq:tvis}).  A disc with a low viscosity in
quiescence diffuses mass slowly to the inner edge. Therefore, the time
to reach $\Sigma_{\rm max}$ is longer when $\alpha_{\rm c}$ is lower
(see top panel of Fig.~\ref{vara}). This also modifies the peak
accretion rate of the outburst.  Similarly, a disc with a higher
viscosity in outburst accretes mass more quickly and the decay time is
shorter (Eqs.~\ref{eq:expop}-\ref{eq:expo}). The conditions for the
disc to enter quiescence and the subsequent evolution are independent
of $\alpha_{\rm h}$.  Therefore, the models shown in the top panel
of Fig.~\ref{vara} which have different $\alpha_{\rm h}$ but the same
$\alpha_{\rm c}$ have the same recurrence time.

\begin{figure}
% makefig8 sur apollo
\resizebox{\hsize}{!}{\includegraphics{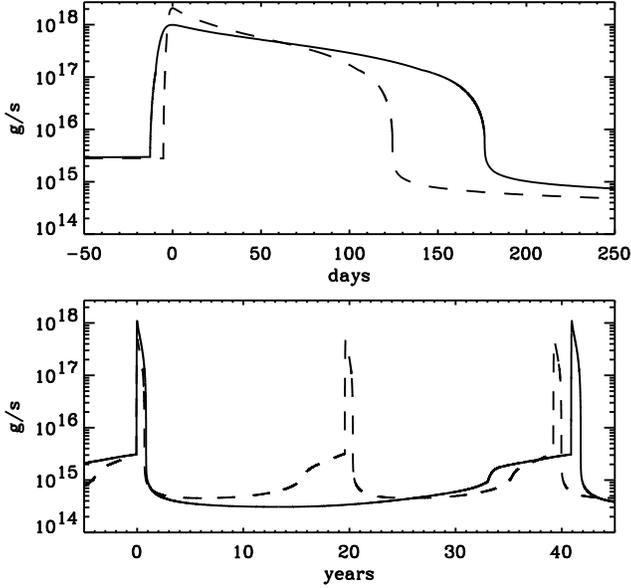}}
\caption{Effect of changing the viscosity parameters on the outburst
time profiles. Bottom: changing $\alpha_{\rm c}$ modifies the
recurrence time (full line $\alpha_{\rm c}=0.01$, dashed line
$\alpha_{\rm c}=0.02$, both with $\alpha_{\rm h}=0.1$). Top: changing
$\alpha_{\rm h}$ modifies the outburst time evolution (full
line $\alpha_{\rm h}=0.2$, dashed line $\alpha_{\rm h}=0.4$, both with
$\alpha_{\rm c}=0.02$. The recurrence time is the same $t_{\rm
rec}\approx 20$~years). See \S\ref{p:vara} for details.}
\label{vara} 
\end{figure} 

\subsection{Changing the mass transfer rate $\dot{M}_{\rm tr}$ \label{p:varmd}}

Next, we explore the effect of changing the mass transfer rate
$\dot{M}_{\rm tr}$, keeping all other parameters identical. For
inside-out outbursts, the time spent in quiescence with fixed $R_{\rm
in}$ is the diffusion time of matter from the outer to the inner edge,
which does not depend on $\dot{M}_{\rm tr}$. However, models including
evaporation show a dependence on $\dot{M}_{\rm tr}$ when $\dot{M}_{\rm
in}$ in quiescence is not negligible compared to $\dot{M}_{\rm tr}$
(Eq.~\ref{eq:trec}).  Fig.~\ref{varmd} shows the quiescence time
shortens with increasing $\dot{M}_{\rm tr}$.

Higher $\dot{M}_{\rm tr}$ result in longer outbursts since (i) the
companion provides more material during the outburst and (ii) the
total disc mass at the onset of the outburst increases with
$\dot{M}_{\rm tr}$. The conditions in the disc are
the same at the end of the outburst, regardless of the mass transfer
rate (except, of course, at the outer edge) because cooling happens in
the same way for discs of the same size. A ring at a given radius can
only cool when $T_{\rm irr}$ reaches $10^4$~K and the surface density
behind the front is roughly the same in all cases. Thus, even with
different $\dot{M}_{\rm tr}$, the discs have almost the same total
mass as they enter quiescence (Fig.~\ref{varmd}).

In a truncated disc the drift of material in quiescence will
change with $\dot{M}_{\rm tr}$ because the distance it has to
travel is rather short. Therefore, the conditions at the beginning of
the outburst depend on the mass-transfer rate. In particular, the
ignition radius increases with $\dot{M}_{\rm tr}$ as could be expected:
a larger amount of mass from the outer edge diffuses down the disc and
can trigger the outburst at a larger radius (but as mentioned earlier
outbursts starting at the outer disc edge do not occur). This also
increases the total amount of mass in the disc at the onset of the
outburst and hence the peak accretion rate (see \S\ref{p:rise}).

For the high mass transfer rates the disc stays fully ionized for a
significant fraction of the outburst, leading to a viscous decay.  Our
model with the highest $\dot{M}_{\rm tr}$ show this is accompanied by
significant variations of the outer radius which produce deviations
from the expected exponential (lower panel of Fig.~\ref{varmd}).

\begin{figure}
%makefig9
\resizebox{\hsize}{!}{\includegraphics{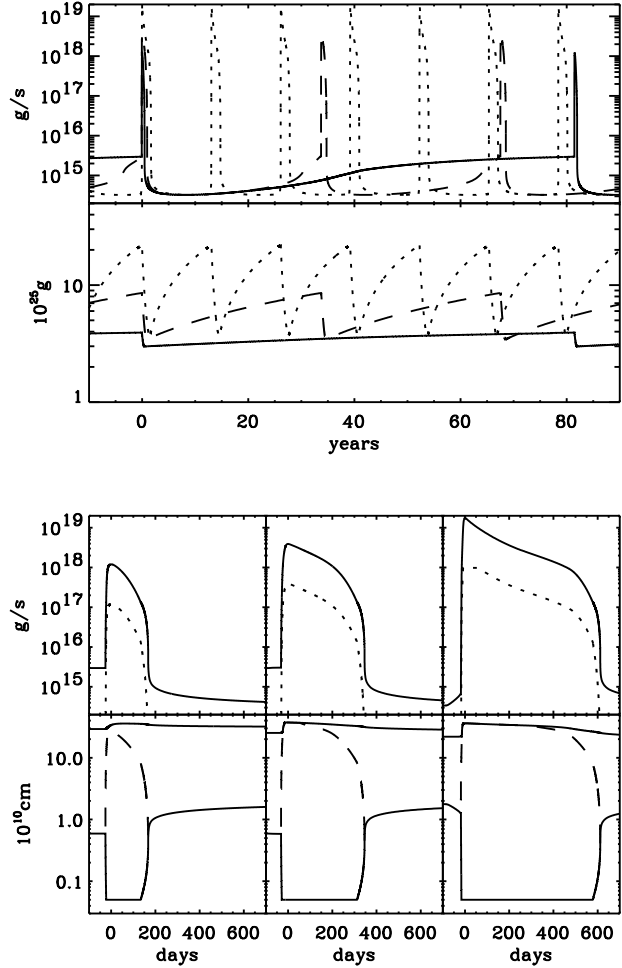}}
\caption{Effect of changing $\dot{M}_{\rm tr}$ on the outburst
cycles. Top two panels : $\dot{M}_{\rm in}$ and $M_{\rm disc}$ for
$\dot{M}_{\rm tr}/10^{16}\mathrm{\ g}\;\mathrm{s}^{-1}=$, 0.5 (full
line), 5 (dashed line) and 50 (dotted line). The bottom panels zoom on
the outburst for the same models, showing $\dot{M}_{\rm in}$ (full
line) and $\dot{M}_{\rm irr}$ (dotted line), $R_{\rm out}$, $R_{\rm
in}$ and the transition radius between the hot and cold regions
(dashed line). The other parameters of the models are $M_1=7$~\msol,
$R_{\rm out}\approx2.5\times 10^{11}$~cm, $\alpha_{\rm h}=0.2$,
$\alpha_{\rm c}=0.02$ and ${\cal C}=5\times 10^{-3}$.  See
\S\ref{p:varmd} for details.}
\label{varmd} 
\end{figure}

\subsection{Changing the disc size $R_{\rm out}$\label{p:varrd}}

The disc size in the model depends on the strength of the tidal
truncation term in the angular momentum conservation equation and on
the circularization radius. The ratios $R_{\rm circ}/a$ and $R_{\rm
out}/a$ depend only on the mass ratio $q=M_2/M_1$. Assuming $q\approx
0.1$, which is reasonable for black hole SXTs, we derive $R_{\rm
circ}/R_{\rm out} \approx 0.5$ (Papaloizou \& Pringle
\cite{pappri2}). In the following, we changed the mean outer disc
radius, keeping the above ratio constant.  This is equivalent to
simulating binaries with the same components but increasing orbital
period $P_{\rm orb}\approx 4.6\ (R_{\rm out}/10^{11}{\rm ~cm})^{3/2}
M_1^{-1/2}$~hours ($q=0.1$).  
%(The 3:1 resonance when
%$R_{\rm out}/a \approx 0.5$ are not taken into account.)
 
\begin{figure}
%makefig10 sur apollo
\resizebox{\hsize}{!}{\includegraphics{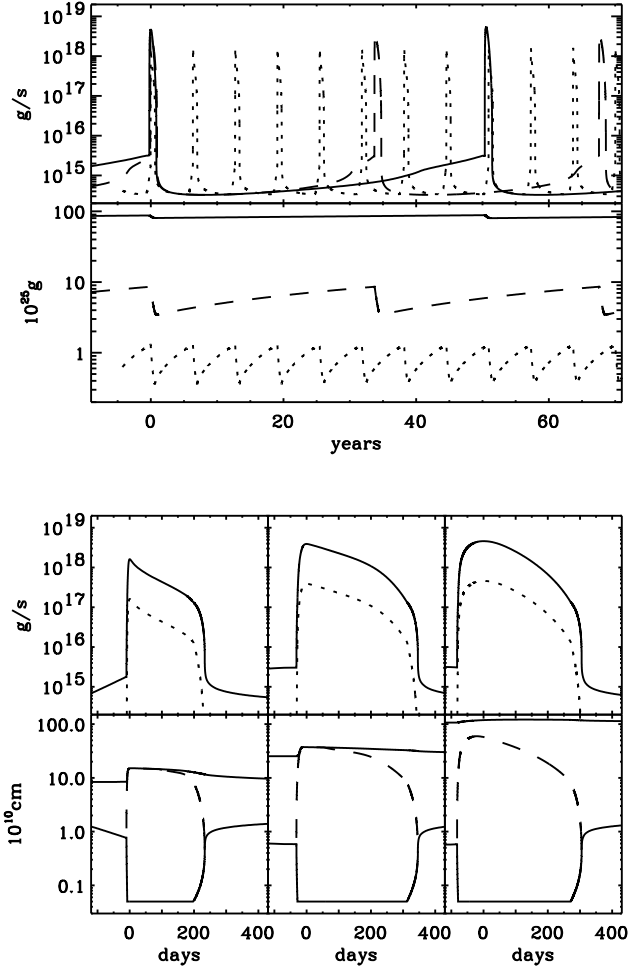}}
\caption{Effect of changing $R_{\rm out}$ on the outburst cycles. Top
two panels : $\dot{M}_{\rm in}$ and $M_{\rm disc}$ for $<$$R_{\rm
out}$$>$$/10^{10}{\rm ~cm}=$ 9.3 (dotted line), 25.7 (dashed line), 107
(full line). The bottom panels zoom on the outburst for the same
models, showing $\dot{M}_{\rm in}$ (full line) and $\dot{M}_{\rm irr}$
(dotted line), $R_{\rm out}$, $R_{\rm in}$ and the transition radius
between the hot and cold regions (dashed line). The other parameters of
the models are $M_1=7$~\msol, $\dot{M}=5\times
10^{16}$~g$\;$s$^{-1}$, $\alpha_{\rm h}=0.2$, $\alpha_{\rm c}=0.02$
and ${\cal C}=5\times 10^{-3}$.  See \S\ref{p:varrd} for details.}
\label{varrd} 
\end{figure} 

Increasing the disc size makes it more difficult for the heat front to
reach the outer disc radius and fully ionize the disc. In small discs
the decay is entirely viscous while larger discs show only
irradiation-controlled linear decays (Fig.~\ref{varrd}). In the case
of a viscous exponential decay the decay rate increases with smaller 
disc radius in accordance with
Eqs.~\ref{eq:expop}-\ref{eq:expo}.

The ignition radius of the outburst increases for smaller disc sizes
since the diffusion timescale is shorter. Smaller discs are also less
massive and, therefore, take less time to replenish than large discs
for the same mass transfer rate: the time spent in quiescence is
shorter. This also leads to smaller outburst peaks (\S\ref{p:rise}).

\subsection{Changing $M_1$ \label{p:varm1}} 

Finally, we vary the mass of the accreting object $M_1$.  Irradiation
is implicitely assumed to be the same since the prescription for
$T_{\rm irr}$ (Eq.~\ref{ill}) does not depend on $M_1$. $\Sigma_{\rm
max}$ and $\Sigma_{\rm min}$ have some dependence on $M_1$
(Eq.~\ref{simax} and \ref{simin}) but their influence on the outburst
is negligible compared to the changes induced by evaporation. The
strength of evaporation depends strongly on $M_1$ through
$\dot{M}_{\rm Edd}$ and $R_S$ in Eq.~\ref{evapor}: $\dot{M}_{\rm
ev}\propto M_1^3$ (this cannot be seen in Menou et al. 2000 where by
error the evaporation formula for neutron-star SXT outbursts uses
$M_1=$ 6 M$_{\odot}$). In outburst, this modifies the accretion rate
at which evaporation sets in which in turn reduces the length of the
outburst for lower $M_1$ (see also Meyer \& Meyer-Hofmeister 2000
who studied extensively the $M_1$ dependence of the instability cycle
with evaporation, but without irradiation, and reached conclusions
similar to those of Menou et al. 1999).

In quiescence, the disc around a lower mass compact object 
is truncated at smaller radii which reduces the
inner accretion rate $\dot{M}_{\rm in}$. Mass is accumulated more
easily (Eq.~\ref{eq:trec}), leading to shorter recurrence time
(Fig.~\ref{varm1}). The lower accretion rates in the evaporated disc
in quiescence do not necessarily translate into lower quiescent
luminosities. In our assumptions, the increase in accretion efficiency
implied by a lower $R_{\rm in}$ ($\epsilon \propto R_{\rm in}^{-2}$)
compensates for the lower $\dot{M}_{\rm in}$. Plotting $\dot{M}_{\rm
irr}=\epsilon \dot{M}_{\rm in}$ instead of only $\dot{M}_{\rm in}$ as
in Fig.~\ref{varm1}, we find $\dot{M}_{\rm irr}\approx
10^{10-11}$~g$\;$s$^{-1}$ for both $M_1=1.4$\msol\ and 7\msol. The
actual efficiency of a 1.4 $M_\odot$ neutron star may be much higher
in quiescence than what we assume here. It might actually be more
correct to assume $\epsilon$ is always 0.1 for the neutron star since
the accreted matter will radiate at the surface. In this case, the
luminosity in quiescence deduced from $\dot{M}_{\rm in}$ will be
higher than for a black hole primary. However, a propeller effect might
significantly reduce the actual mass accretion rate onto a neutron star in
quiescence (Menou et al. 1999c) making the distinction between black holes
and neutron stars less obvious.

\begin{figure}
%makefig11 sur apollo
\resizebox{\hsize}{!}{\includegraphics{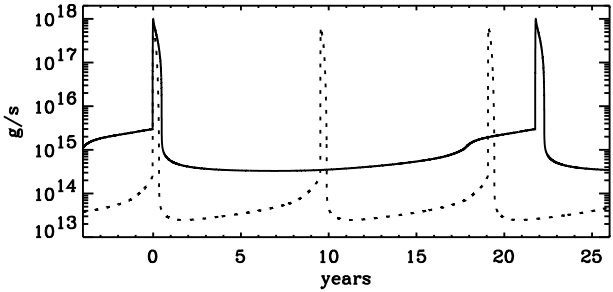}}
\caption{Effect of changing $M_1$ on the outburst cycles of the
inner mass accretion rate $\dot{M}_{\rm in}$. The model of \S5 is
shown by the full line ($M_1=7$~\msol). The same model but with
$M_1=1.4$~\msol\ is shown by the dotted line. See \S\ref{p:varm1} for
details.}
\label{varm1} 
\end{figure} 

\section{Discussion}

Our most basic hypothesis is that the outbursts of SXTs are due to the
thermal-viscous instability in a thin accretion disc. There is little
doubt that LMXBs do possess thin accretion discs at least far from the
compact object, while the wide applicability of the thermal-viscous
instability in CVs gives strong support to its existence in LMXBs. It
has long been known that models of dwarf novae require a lower
viscosity in quiescence than in outburst. Models with $\alpha_{\rm
c}=\alpha_{\rm h}$ predict low-amplitude outbursts with short cycles
which do not agree with observations (Smak \cite{smak84}; Lin et
al. \cite{lpf}).  The upper hot branch of the local thermal
equilibrium curve (\bS-curve) is characterized by a high viscosity
while the opposite is true for the lower cold branch. Our 
`standard' model refers to the thermal-viscous instability in
such a two-$\alpha$ thin disc.

These two-$\alpha$ models, although not without problems (Smak
\cite{smak2000}; Lasota \cite{review}), have basic properties
(recurrence time, amplitude and length of some type of outbursts)
which compare reasonably well with observations, at least for
some classes of CVs (for instance, the U Gem type dwarf novae). The
standard model as applied to CVs offers a background upon which more
detailed scenarios have been constructed (with various success) to
explain observational features such as the UV delay, superoutbursts,
superhumps, standstills etc. (see Lasota \cite{review} for a
review). None of these features have (up to now) threatened the basic
premise that the outbursts are triggered by the thermal-viscous
instability in a two-$\alpha$ thin disc.

In contrast, the standard DIM was {\em never} able to reproduce even
the most basic features of {\em any} SXT.  In particular, the
impossibility of obtaining anything like the observed exponential
decays casted serious doubts on the validity of using the two-$\alpha$
assumption to represent the viscosity in these discs, prompting the
use of more complex formulations (Cannizzo et al.  \cite{ccl}). After
all, disc models with {\em ab initio} treatment of the viscosity are
still some way off and the $\alpha$-disc could be an
oversimplification.  However, one must ensure that the failure of
models to reproduce observations are really due to the viscosity
prescription, and not to the fact that some physical effects have not
been correctly taken into account.

Indeed, in a number of cases it has been shown that the disagreements
between models and observations resulted not so much from the
assumptions on the viscosity but on the physics which was or was not
included. In CVs, studies that successively included a variable disc
size, evaporation of the central regions, irradiation by the white
dwarf, stream-impact and tidal dissipation heating have gradually led
to increased improvement in the agreement between observations and
detailed predictions of models (the most recent being
Hameury et al. \cite{hameury}; \cite{hameury2}; \cite{hameury4};
Buat-M\'enard et al. \cite{bhl1}). For example, it was shown by Smak
(1998) that the failure of models to reproduce the observed UV delay
largely resulted from the use of a fixed outer boundary condition 
and too small discs, making outside-in outbursts quite difficult to
obtain. And of course, it was early realized that parameters such as
the mass transfer rate from the secondary had to vary in order to
account for the standstills of the Z Cam stars (Meyer \&
Meyer-Hofmeister 1983), not to mention the necessity of using a number
of grid points large enough to resolve fronts (Cannizzo 1993).  It
will therefore come as no surprise that the failure of the
two-$\alpha$ model in SXTs was more indicative of missing physics than
inadequate assumptions on the viscosity.

In the following we review the results of the previous sections to
show how the fundamental properties of evaporation and irradiation
lead to better agreement with observations within the two-$\alpha$
model.  We then critically examine possible failures of this model in
light of detailed comparisons with observations.

\subsection{Fundamental properties of the model}

It has long been known that the standard disc instability model fails
on three fundamental accounts: (a) the timescales of the cycles are
wrong with short outbursts and recurrence times; (b) the decay from
outburst is dominated by reflare episodes which are not observed; (c)
even the full conversion to X-rays of the rest mass energy of the
accreted material could not power the observed luminosities if the
quiescent disc reaches down to the compact object.  All of these are
unavoidable consequences of the model which do not depend on the
detailed assumptions. In the standard DIM thermal cooling starts as
soon as the outburst peak is reached.  This proceeds on a short
timescale since an unstable ring evolves on a thermal
timescale. This results in (a). Problem (b) is the consequence of the
high densities behind the cooling front and of the decreasing critical
density needed to trigger an outburst as the cooling front propagates
inwards. Unless the disc can be emptied in outburst, which is
difficult because of (a), reflares are unavoidable. Problem (c) is
also due to the decreasing critical density: a quiescent disc must
have densities below the critical $\Sigma_{\rm max}$ which implies
extremely low accretion rates if the disc extends to small
radii. Either the quiescent X-rays are not produced by accretion
(which is possible for neutron stars: Rutledge et al. \cite{rutledge},
but very unlikely for black-holes: Lasota \cite{lasota2}) or the disc
is truncated so that the mass flow through the inner region (assuming
it is constant) is high. This was first suggested in the context of
dwarf novae to explain the optical to EUV delay (Meyer \&
Meyer-Hofmeister \cite{mmh}).

This is a particularly attractive scenario when the inner region is a
hot ADAF flow (Lasota et al. \cite{lny}). An ADAF of varying size and
accretion rate produces harder or softer spectra which compare well
with the observed X-ray spectral states of SXTs. With this in mind,
Menou et al. (\cite{menou1}) included disc truncation in quiescence to
the standard DIM. Yet, their models required values of $\alpha$ lower
than those usually assumed in dwarf novae and were not able to
reproduce the short rise times observed in SXTs. In principle one
would hope the viscosity behaves in the same way between CVs and
LMXBs\footnote{But one should keep in mind that there is a subclass of
dwarf novae (the WZ Sge systems) for which low values of the viscosity
in quiescence have been invoked}. Another way to get long recurrence
times would be to have the rate of evaporation of the disc in
quiescence almost match the mass transfer rate from the secondary
(Meyer \& Meyer-Hofmeister \cite{meyer4}). Again this is not
satisfactory: this requires fine-tuning of the evaporation process
and/or strongly implies that most of the LMXB population would
be cold and stable (Menou et al.  \cite{menou3}; Meyer \&
Meyer-Hofmeister \cite{meyer3}) with the outbursts triggered by some
unspecified mechanism (e.g. increased mass transfer from the
secondary).  This is not unlikely, but is disappointing in that most
SXTs would then be unusual in some respect with little prospect for
unifying explanations.

Irradiation provides an adequate resolution of the three problems
mentioned without having to assume a very low quiescent viscosity.
Irradiation is an observational fact and simple estimates show its
impact on the outburst cycles must be important.  Irradiation is also
the only major difference between CVs and LMXBs. In CVs the
irradiation flux is at lower energies and originates from a
(relatively) large white dwarf. In LMXBs the flux is at high energies
and (presumably) originates from the central regions of the disc or
from a corona. The distribution and amplitude of the fluxes are
totally different. As such it was a prime candidate for the most
important missing piece of physics in SXT models.

In \S4 we have seen how irradiation drastically changes the outburst.
The key point is that the thermal instability is suppressed when
$T_{\rm irr}\gta 10^4$~K. In this sense, the results shown in this paper
depend little on the assumptions made on the irradiation flux. A disc
will show slow exponential decays for as long as a large enough
irradiating flux can keep the outer disc above hydrogen ionization and
the exact formulation of $F_{\rm irr}$ is unimportant. The first major
consequence is therefore to prolong the outburst by providing a
significant additional source of heating. 

During the thermal decay, the irradiation flux determines the location
of the cooling front. In general the cooling front can only appear
after most of the mass of the disc has already been accreted. Even if
irradiation stops too quickly and the outburst ends prematurely, the
densities in the disc will still be much lower than when irradiation
is not included. This is our second major consequence since this makes
reflares impossible and also lengthens the quiescence time needed to
rebuild the disc mass.

In our models we have neglected the possible impact of the X-ray
spectrum on irradiation heating. We have implicitly assumed it was
soft enough to confine heating to a small layer at the top of the disc
(Lyutyi \& Sunyaev \cite{lyutyi}). On the other hand, hard X-rays
should penetrate deeper and deposit energy all along the vertical
layer. This should not significantly change the main conclusions of
this paper for a simple reason: whatever the distribution of
irradiation heating in the layer it will still globally heat the ring
and displace its thermal equilibrium \bS-curve to lower $\Sigma$. This
can be seen in Figs.1-2 of El-Khoury \& Wickramasinghe
(\cite{ekw}). Even hard X-rays will prolong the outburst until
sufficiently low densities can be reached for the disc to cool.

In an irradiated disc extending to low radii problem (c) remains. By
including truncation in \S5, we have shown that very long recurrence
times are easily obtained. These result naturally from irradiation
which forces the disc to accrete more during outburst and from
evaporation which requires the disc to reach higher densities before
an outburst can be triggered again. There is no need for very low
values of $\alpha$ in quiescence to reach recurrence timescales of tens
of years.

An adequate model of SXTs can therefore be obtained by including some
form of irradiation in outburst and some form of evaporation in
quiescence. Regardless of the detailed assumptions this will
necessarily lead to long outbursts with slow decays and long
quiescence times for values of the viscosity which are similar to
those used in models of CVs.

\subsection{Comparison with observations}

The observed timescales for the rise (1-14 days), decay (8-40 days),
and total outburst duration (40-300 days) defined by Chen et al. 
(\cite{chen}) are well reproduced by our models in very general
conditions i.e. with no fine tuning of any parameter, including
$\alpha$. The model also predicts recurrence timescales anywhere
between a year to tens of years, as observed. The amplitudes of the
outbursts are high with peaks at $10^{37-38}$~erg$\;$s$^{-1}$
(assuming $L_{\sc x}=0.1 \dot{M}_{\rm peak} c^2$), all consistent with
observations. Weak outbursts (i.e. with less irradiation) have shorter
durations than strong ones just as noticed by Chen et al. (\cite{chen}). 

Our models, taken at face value, do not make accurate
predictions for the observed outburst amplitudes in neutron stars and
black holes because the efficiency $\epsilon$ of the ADAF in
quiescence is not properly calculated here. It would be particularly
interesting to see whether quiescent discs with neutron star primaries
have similar luminosities as black hole systems as briefly suggested
in \S\ref{p:varm1}. In this case, this would support the idea that the
higher quiescent luminosities of NS SXTs is due to thermal emission
from the surface of the NS heated by accretion during the outburst and
hence proves indirectly the existence of horizons in BHs (Narayan
et al. \cite{narayan}; Menou et al. \cite{menou4})

A clear feature of the model which is observed in SXTs is the slower
optical decay time. This is due to irradiation which, under very
general conditions, can keep the cold outer disc hotter than usual
during the propagation of the cooling front
(Fig.~\ref{fig:optical}). Fig.~\ref{fig:decaytidim} shows the disc
behind the front is dominated by irradiation and almost isothermal.
Obviously, the peak optical flux is much higher when the disc is
irradiated than when it is not.

\begin{figure}
%makefigdelay sur apollo
\resizebox{\hsize}{!}{\includegraphics{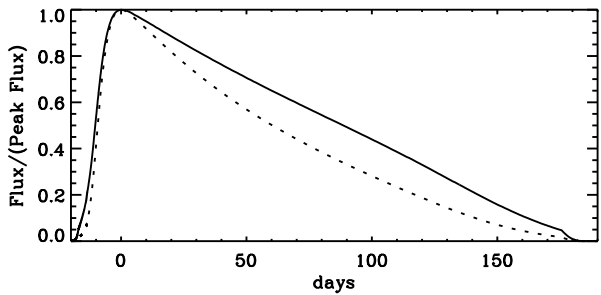}}
\caption{V-band optical flux (full line) compared to $\dot{M}_{\rm
irr}=\epsilon \dot{M}_{\rm in}$ (dotted line) during an outburst of
the model of \S5 (Fig.~\ref{fig:tidim}). $\dot{M}_{\rm irr}$ is taken
as an indicator of the X-ray flux. Both are normalized to their peak
value. The optical flux is halved in $\approx 100$~days while the
X-ray flux is halved in $\approx 60$~days.}
\label{fig:optical}
\end{figure}

There is observational evidence pointing towards a delay between the
outburst rise in the optical and the rise in the X-rays. This has been
most evident in GRO J1655-40 where a 6 day delay was observed (Orosz
et al. 1997). As stressed by Hameury et al. \cite{hameury3}, this delay and
the light-curve shape cannot be explained by an outside-in
outburst in a disc extending down to the last stable orbit. The delay
is rather the time for the thin disc to replace the hot flow in the
inner regions. This happens on a much longer viscous timescale than
the timescales on which fronts propagate through the disc. In the
early rise, one would expect the optical to rise first due to the
propagation of the heat front and the X-rays to form later when the
thin disc reaches regions close to the compact object. Whether the
outburst starts with an inside-out or an outside-in front matters
little.

In the present calculations, the time taken by the thin disc
to reach its minimum radius is of the order of a few days (see
\S\ref{p:outburst}).  Although it would be desirable, we cannot
provide better tests of the delays as this would require a detailed
model of X-ray formation and of irradiation during the early rise
(which can drastically change the optical lightcurve).  This can be
done if one couples an ADAF model with the thin disc and consistently
calculates the evolution of both, specifically the thin disc
irradiation from the ADAF.

To zero order the model thus satisfies all major observed
properties.  Inversely, what can be inferred from an observed
lightcurve on the accretion disc ?  An easily obtained measure is the
peak flux of the outburst.  The study of \S6 shows this increases with
mass accretion rate and disc size (orbital period). The peak
luminosity for neutron star primaries is generally smaller than for a
black hole with a similar accretion disc. However, this might be
dependent on the assumed evaporation law (here evaporation is weaker
for a NS). Another parameter of interest, when available, is the
recurrence time between outbursts. The models predicts the lowest
times for systems closest to the stability limit i.e. with high mass
transfer rates for their sizes (or vice-versa). Systems with long
orbital periods are more likely to show outbursts deviating from an
exponential since the heat front may not reach the outer edge.  The
strength of irradiation is also important : systems showing little
evidence for it would have short outbursts. The above assumes identical
values of the viscosities. In practice, the effects of the different
parameters are too entangled to reach strong conclusions from an
observed lightcurve except, perhaps, in the most extreme cases.
Consequently, the prospects of further constraining viscosity from the
observed outburst lightcurves are {\em dim}. Dwarf novae, if only
because of sheer numbers, are a better tool to achieve this goal
(e.g. Smak \cite{smak99}).  X-ray observations of SXTs in quiescence
will also be very helpful in constraining both the evaporation
processes and the viscosity at low temperatures, since very low
quiescent viscosities should result in very small X-ray luminosities.

\subsection{Outstanding problems}

A quick review of observed X-ray lightcurves will reveal that
these are much more complex than the nicely behaved decays we
predict. Indeed, the past few years have seen a number of SXTs
detected with lightcurves which seem destined to ruin a model
originally based on the canonical \object{A0620-00} (e.g. \object{GRO
J1655-40}, \object{XTE J1550-564}, \object{XTE J1118+480}...). These
systems typically show consecutive outbursts separated by a few months
with the second outburst more akin to a plateau than an exponential.
Both the plateau and short quiescence time (if the system returns to
quiescence) cannot be explained by the model presented here without
additional assumptions. Possibilities include increased mass transfer
from the secondary after the star was irradiated during the first
outburst. This can easily lead to plateau-like lightcurves when mass
transfer temporarily stabilizes the disc.  Further complications can
arise if the disc is large enough that it becomes warped in outburst
and, for instance, shadows the secondary (Esin et al. \cite{esin};
Ogilvie \& Dubus \cite{ogilvie}). Winds created by strong irradiation
can cause significant mass loss resulting in quicker exponential
decays (Mineshige et al. \cite{myi}).

The most decently behaved SXTs also have some intriguing features.
For instance, \object{A0620-00} and \object{GRS 1124-68} show
glitches (in the terminology of Chen et al. \cite{chen})
during a decay which otherwise could very well be fitted by the model
(see the discussion by Cannizzo \cite{cannizzo3}). Our models clearly
show these cannot be due to the transition between the viscous and
thermal decays as speculated by King \& Ritter (\cite{kr}).  Esin et
al. (\cite{esin3}) proposed that the glitches could be due to
increased irradiation in a warped disc while Cannizzo
(\cite{cannizzo4}) argued for evaporation in the outer disc. Both
these possibilities have been explored and will be dealt with in a
future paper. Even more puzzling are the multiple `mini-outbursts'
observed at the end of the outburst of e.g. \object{GRO
J0422+32}. These bear many similarities to some outbursts of WZ Sge
type dwarf novae (e.g. Kuulkers \cite{kuulkers}), suggesting a common
mechanism.  The mini-outbursts are in no way comparable to the
reflares discussed above. There are no satisfying explanations for the
mini-outbursts.

\section{Conclusion}

We have shown that, under general assumptions, a model of the thermal
viscous instability including both the effects of irradiation and
evaporation of the inner disc in quiescence led to outburst cycles
with timescales, amplitudes and slow decays which compare well with
observations.  This does not require any fine-tuning or modifications
of the standard viscosity assumptions used in models of
CVs. Even very moderate amounts of irradiation significantly increase
the length of the outburst, its shape and the recurrence timescale.

These results depend little on the detailed choices made in describing
either the irradiation flux or the evaporation law. The assumptions
are most important during the rise and at the end of the
outburst. Further comparisons with observations will require these to
be known more precisely.  The next logical step will be to
consistently include an ADAF model in the evaporated region so as to
be able to calculate at each moment the X-ray spectrum and flux seen
by the disc (and the observer).

The wide applicability of the DIM in CVs and SXTs is therefore
confirmed. The importance of a careful assessment of the necessary
physics cannot be more stressed than for the case of SXTs whose basic
features can only be reproduced by the model when the effects of
irradiation and evaporation are included.

Yet, the lightcurves of SXTs show more complexity than the simple
exponential or slow decays we predict. This is suggestive of still
more missing physics, possibly variations in the mass transfer rate
from the secondary or complex behaviour of the irradiation flux. These
are difficult to model and most certainly vary from object to
object. In this sense, more detailed modelling might not prove very
rewarding. Also the prospects of using SXTs to
constrain viscosity beyond the simple two-$\alpha$ model are poor. The finer
effects of viscosity are hidden amongst the complex interactions
between other physical effects.  However, the dynamical models
show viscosity must behave in the same way in CVs and SXTs which, at
least, supports a common origin.

\section*{Acknowlegements}
We thank Hans Ritter for his very detailed report.  GD
is supported by the TMR network `Accretion on to Black Holes, Compact
Stars and Protostars' (contract number ERBFMRX-CT98-0195). This work
was also supported by a Spinoza award to E.P.J. van den Heuvel and by
the {\sl Programme National de Physique Stellaire} of the CNRS.

\listofobjects
\end{document}